\documentclass[prb,reprint,sortedaddress,10 pt]{revtex4-1}

\usepackage{graphics}
\usepackage{graphicx}
\usepackage{epsfig}
\usepackage{subfigure}
\usepackage{ulem}

\usepackage{amsthm,amsmath,amssymb,amsfonts,bm,subfig}

\usepackage{graphicx}
\usepackage[usenames]{color}
\usepackage{tikz}

\definecolor{MyGray}{rgb}{0.92,0.93,0.94}

\newcommand{\tz} {\hat z}

\newcommand{\ba} {\bm a}

\newcommand{\hpsi} {\psi}

\newcommand{\bx} {{\bm x}}

\newcommand{\bz} {{\bm z}}
\newcommand{\br} {{\bm r}}
\newcommand{\bq} {{\bm q}}
\newcommand{\Q} {{\bm Q}}
\newcommand{\R} {{\bm R}}
\newcommand{\DDQ} {{\bm S}}
\newcommand{\xsi} {{\xi^2}}
\newcommand{\buz} {\underline {\bm z}}

\newcommand{\Diag}[1] {\mbox{diag}\left( #1 \right)}

\begin{document}

\title{Augmented projections for ptychographic imaging}

\begin{abstract}
  Ptychography is a popular technique to achieve diffraction limited
  resolution images of a two or three dimensional sample using high
  frame rate detectors.  We introduce a relaxation of common
  projection algorithms to account for instabilities given by
  intensity and background fluctuations, position errors, or poor
  calibration using multiplexing illumination. This relaxation
  introduces an additional phasing optimization at every step that
  enhances the convergence rate of common projection algorithms.
  Numerical tests exhibit the exact recovery of the object and the
  perturbations when there is high redundancy in the data. 
\end{abstract}

\author{Stefano Marchesini }
\affiliation{Advanced Light Source, Lawrence Berkeley National Laboratory, Berkeley, CA 94720}
\email{smarchesini@lbl.gov}
\author{Chao Yang} 
\affiliation{Computational Research Division,
Lawrence Berkeley National Laboratory, Berkeley, CA 94720.}
\email{cyang@lbl.gov}
\author{Hau-tieng Wu}
\affiliation{Statistics, University of California, Berkeley, CA 94720}
\email{hauwu@berkeley.edu}
\author{Andre Schirotzek}
\affiliation{Advanced Light Source, Lawrence Berkeley National 
  Laboratory, Berkeley, CA 94720}
\email{aschirotzek@lbl.gov}
\author{Filipe Maia}
\affiliation{NERSC, Lawrence Berkeley National Laboratory, Berkeley, CA 94720}
\email{frmaia@lbl.gov}
\maketitle

\begin{figure*}[hbtp]
  \begin{center}
%    \includegraphics[width=0.4\textwidth]{exp1.pdf}
%\ \ \ 
    \includegraphics[width=0.9\textwidth]{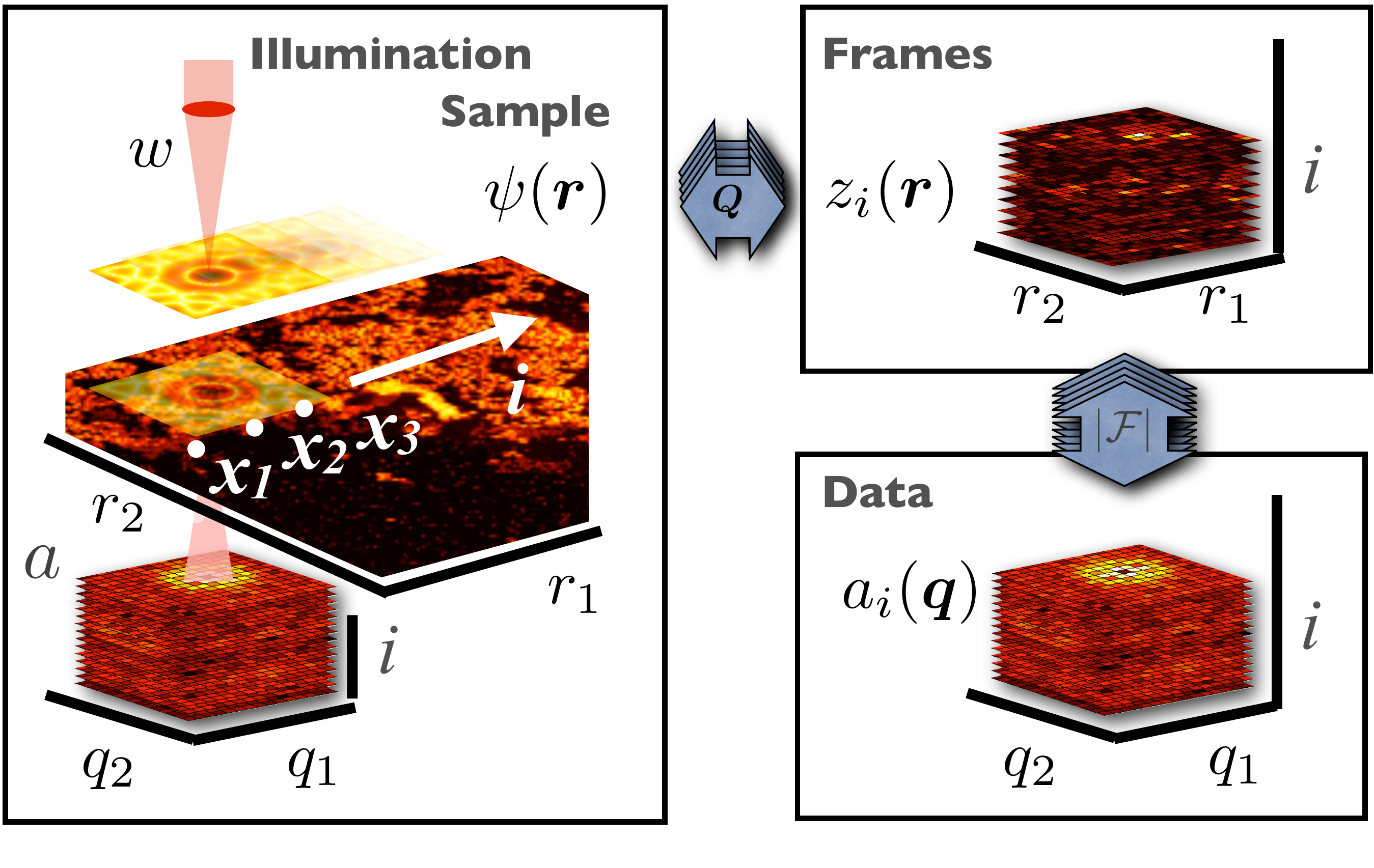}
 \end{center}
 \caption{(Left) Experimental geometry in ptychography: An unknown
   sample with transmission $\psi(\bm r)$ is rastered through an
   illuminating beam $w(\bm r)$, and a sequence of diffraction
   measurements $|a_{(i)}|^2$ are recorded on an area detector as the
   sample is rastered around. The point-wise product between
   illuminating function and sample $z_{(i)}(\bm
   r)=w(\br)\hpsi(\br+\bx_{(i)})$ -which we refer to as ``frame'' throuout
   the paper- is related
to the measurement by a Fourier magnitude relationship $a_{(i)} =
\left |{\cal F} z_{(i)} \right |$.
\label{fig:exp}}
\end{figure*}

\section{Introduction}

Ptychography was proposed in 1969 to improve the resolution in electron or x-ray microscopy
\cite{Hoppe,hegerl,batesptycho, spence:book}. In a scanning microscope, a small beam is focused onto the sample via a lens, and the transmission is measured in a single-element detector. The image is built up by plotting the transmission as a function of the sample position as it is rastered across the beam. In such microscope, the resolution of the image is given by the beam size. In ptychography, one replaces the single element detector with a two dimensional array detector such as a CCD and measures the intensity distribution at many scattering angles. Each recorded diffraction pattern  contains short-spatial Fourier  frequency  information about features that are smaller than the beam-size, enabling  higher resolution. At short wavelengths however it is only possible to measure the intensity of the diffracted light. To reconstruct an image of the object, one needs to retrieve the phase information. With  measured amplitude and phase information, a high resolution image can be readily computed,  phase contrast imaging becomes possible, and the depth of focus for 3D micro-tomography is no longer a problem. 
While  phase retrieval problems are notoriously difficult to solve numerically,  the  problem is made tractable in ptychography by using redundant measurements. In practice, multiple views of the same region of the object are recorded by using a small step size -relative to the size of the illuminating beam- when scanning the sample. With high speed detectors\cite{pilatus,fastccd}  and ever brighter light sources, ptychographic imaging is becoming increasingly popular.

A practical issue in  ptychographic reconstruction  are the strict requirements of the experimental geometry to achieve high quality data. 
 For example, the need for stable, well controlled coherent illumination of the sample, limited detector speed and response function all contribute to limit the specifications of a ptychographic microscope. 
New methods to work with unknown illuminations were proposed\cite{Chapman1996,Thibault2008,thibault:probe, pie2}. They are now used to calibrate high quality x-ray optics  \cite{Kewish2010,Honig:11,guizar-mirror} and space telescopes. More recently,  position errors\cite{fienup,Maiden:2012, axel}, background\cite{thurman2009,guizar:bias}, noise statistics \cite{Thibault-Guizar,Godard:12} and partially coherent illumination\cite{clark:2011,Fienup:93,Abbey2008,CDIpartialcoherence,spence-flip},  or vibrations have been added to the nonlinear optimization to fit the data. 

{

Existing methods iterate between an object space -  an image representing an estimate of the object - and a measurement space given by the measured diffraction frames. 
For example in the approach first described by \cite{fienup}, one starts from an estimate of the positions of the illuminating beam and an estimate of the object under study. From this starting point one  minimizes the discrepancy with the data using local - gradient based  - optimization
and obtains a new estimate of the object and positions. 

In this paper we introduce an additional optimization step in the measurement space 
 which is -since  in ptychographic experiments one records multiple views of the same region of the object  - of higher dimensionality than the object space.  
  We form pairwise comparisons between neighboring frames and update the unknown parameters of each diffraction frame so that each frame is consistent with each other. 

The approach described in this work achieves accelerated convergence for large scale phase retrieval problems spanning multiple  length-scales. We also show that this approach can  recover experimental fluctuations over a large range of time-scales.  
}

\subsection*{Notation and background}

 In a ptychography experiment
\cite{batesptycho,rodenburg,Chapman1996,pie1,pie2,Thibault2008, Dierolf}
(see Fig. \ref{fig:exp}),  a two dimensional small beam with distribution $w(\bm r)$
of dimension $m \times m$ illuminates an subregion centered at $\bx_{(i)}$ -referred to as frame- of an unknown object of interest
$\hpsi$ of dimension $n\times n$. 
Here 
{ $0<m< n$, $i=1,\ldots,k$, $k$ is the number of frames, $r>0$ is a lengthscale
(the diffraction limited resolution of the microscope), and}
\begin{align*}
&\br= \left (r\mu,r\nu \right),\,\, \mu,\nu\in\{0,\hdots, m-1\},\\
&\bx_{(i)}= \left(r\mu',r\nu'\right),\,\,\mu',\nu'\in\{0,\hdots,n-m\}.
\end{align*}

As $\bx_{(i)}$ is
rastered on a typically coarser grid, $\bm r+\bx_{(i)}$ spans a finer grid of dimension $n\times n$.
Here for simplicity we consider square matrices, and generalization to
non-square matrices is straightforward but requires more indices and
complicates notation. 
In other words, we assume that a sequence of
$k$ diffraction intensity patterns ${\cal{I}}_{(i)}(\bq)=a_{(i)}^2(\bq)$ are
collected as the
position of the object is rastered on the position $\bx_{(i)}$, where 
$$
\bq=\left (\tfrac{2 \pi}{m r }\mu,\tfrac{2 \pi}{m r } \nu \right),\,\,\mu,\nu\in\{0,\hdots, m-1\}.
$$
%Each frame $a_{(i)}$ contains $m \times m$ pixels.  
The relationship among the amplitude $a_{(i)}$,  the probe $w$ and an unknown
object $\hpsi$ to be estimated can be expressed as follows:
\begin{eqnarray}
\label{data}
&&a_{(i)}(\bq) = \left |{\cal F} w(\br)\hpsi(\br+\bx_{(i)}) \right |
,\nonumber\\
%&& \br=r \left (\mu,\nu \right),\,\, \bq=\tfrac{2 \pi}{r}\left (\mu,\nu \right),\nonumber\\
%&&\mu,\nu\in\{0,\hdots, m-1\},\nonumber
&&({\cal F} f )(\bm q)= \sum_{\br} e^{i \bq \cdot \br} f(\br),\nonumber
%\\
%\nonumber
%&&\bm m = \left (\mu,\nu \right), 
%\quad
%&& ,
\end{eqnarray}
where 
the sum over $\br$ is given on all the indices $m \times m$ of $\bm r$,
and ${\cal F}$ is the two dimensional discrete Fourier transform.

We introduce the illumination operator $Q_{(i)}$, $i=1,2,\cdots, k$, associated with
 $\bx_{(i)}$ that extracts a frame $z_{(i)}$ out of $\hpsi$, and scales
the frame point-wise by the illumination function $w(\bm r)$ (see Fig.
\ref{fig:exp}):
\begin{eqnarray}\nonumber
&&
z_{(i)}(\br)= w(\br) \hpsi(\br+\bx_{(i)})=[Q_{(i)} \hpsi](\br),
\nonumber
%&&Q_{(i)}(\br)=w(\br) e^{i \bx_{(i)} \partial_\br}.
\end{eqnarray}
where $z_{(i)}$  represents the frames extracted from $\psi$ and
multiplied by the probe $w(\br)$.

To have a compact representation for numerics, we introduce the
following notations. We represent $\hpsi$ as a vector of length $n^2$,
that is, $\hpsi\in \mathbb{C}^{n^2}$.  The moving beam associated with
the illumination function $w(\bm r)$ can be represented as an $m^2
\times n^2$ sparse ``illumination matrix'' associated with the
illumination operator, which is again denoted as $Q_{(i)}$.  To
express $Q_{(i)}$ in the matrix form, we introduce a restriction
matrix $R_{(i)}$, which restricts the $n\times n$ region onto the
$m\times m$ subregion centered at $\bx_{(i)}$, that is,
 $$
 Q_{(i)}=\text{diag}(w)R_{(i)}.
 $$

The relationship between the diffraction measurements collected in a
ptychography experiment and the unknown object to be recovered can be
represented compactly as:
\begin{eqnarray}
\ba&=&|{\bm F} \Q  \hpsi|,
\label{eq:data}
\end{eqnarray}
if we stack the diffraction measurements $a_{(i)}$ into a long
vector $\ba$, and define various matrices as follows:
\begin{eqnarray}
\nonumber
\ba&=&
\left(
\begin{array}{c}
a_{(1)} \\
\vdots \\
a_{(k)}
\end{array}
\right),\,
\Q=
\left(
\begin{array}{c}
Q_{(1)} \\
\vdots \\
Q_{(k)}
\end{array}
\right)
,\,\\
\nonumber
 \bz&=&
\left(
\begin{array}{c}
 z_{(1)} \\
\vdots \\
 z_{(k)}
\end{array}
\right),
\,
\bm F=
\left(
\begin{array}{ccc}
{\cal F}\\
 & \ddots &  \\
 &  &  {\cal F}
\end{array}
\right)\,.
\end{eqnarray}
We call the domain of $\Q$ the {\it object space}, and the range of 
 $\bm F\bm Q$ the {\it measurement space}.

Geometrically, $\Q$ is the matrix that extracts $k$ frames out of an
object $\psi$ and multiplies them by the probe $w$; $\Q^\ast$ is the
conjugate transpose that merges $k$ frames onto the object space; in
addition, $\Q^\ast\Q$ can be viewed as the normalization factor given
by the sum of the illumination functions. In particular, by a direct
calculation, $\Q^\ast\Q$ is a $n\times n$ diagonal matrix whose $l$-th
diagonal entry is 
{$\sum_{k:\, l=\bx_k+\br_k }|w(\br_k)|^2$}, where we
abuse the notation by using $l$ to indicate the point on the object
space.  
{ Physically $l$ is the index of the grid point on the unknown object of interest which is covered by the point $\br_k$ of the $k$-th illumination window.}
  See table \ref{tab:algebra} for the relationship
  between probe $w$, translation $\bx_{(i)}$, $Q_{(i)}$ and $\Q$.

\begin{table}[htbp]
  \center
  \begin{tabular}{|l|l|}
    \hline
$Q_{(i)} \psi$       & $\text{diag}(w)R_{(i)}\psi$ 
\\ 
\hline 
$Q_{(i)}^\ast z_{(i)}$ & $ R_{(i)}^*\text{diag}(\mathrm{conj}(w)) z_{(i)}$ 
\\ 
\hline
$\Q^\ast \bz$      & $\sum_{i} R_{(i)}^*\text{diag}(\mathrm{conj}(w)) z_{(i)}$ 
\\ 
\hline
$e_l^T\Q^\ast\Q e_k$        &  $\sum_{\bx_k+\br_k=l}|w(\br_k)|^2\delta_{l,k}$ %\ldots \sum_{(k,\br_k)=n^2}|w(\br_k)|^2])$ %$\left\{\begin{array}{lll} \sum_{\br_k}|w(\br_k)|^2 && (l,l)\mbox{-entry}\\ 0 && \mbox{o.w.} \end{array} \right.$     
\\ 
\hline
$e_l^T(\Q^\ast\Q)^{-1} \Q^\ast \bz $ & $\frac{e_l^T\sum_{i} R_{(i)}^*\text{diag}(\mathrm{conj}(w)) z_{(i)}}{\sum_{\bx_k+\br_k=l}|w(\br_k)|^2}$ 
\\ 
\hline
$P_{\bm F}^{\bm a_{(i)}} z_{(i)} $ & $\sum_{\bq} e^{-i \bq \cdot \br} \frac{\sum_{\br} e^{i \bq \cdot \br}
  z_{(i)}(\br)}{|\sum_{\br} e^{i \bq \cdot \br} z_{(i)}(\br)|} \bm{a}_{(i)}$
\\ 
\hline
\end{tabular}
\caption{
  Linear algebra notation. Here $e_l$ is the unit 
$n\times 1$ vector with the $l$-th entry $1$ and
 $\delta$ is the Kronecker delta.
 The division is understood as an element-wise operation.
 {The operator $P_{\bm F}^{\bm a_{(i)}}$ is defined in (\ref{projection_f})}
 \label{tab:algebra}
}\end{table}

The objective of the ptychographic reconstruction problem is to find
$\hpsi$ given $\ba$  from (Eq. \ref{eq:data}).  This is often formulated using a ``divide and conquer'' approach referred to as \textit{projection algorithms},
\textit{iterative transform methods}, or \textit{alternating direction
  methods} \cite{ZWen}.  
One formulates the
relationship (Eq. \ref{eq:data})  as:

\begin{eqnarray}
\ba&=&|{\bm F} \bz|,
\label{eq:data_F}
\\
\bz&=&\Q\hpsi.
\label{eq:overlap}
\end{eqnarray}

These algorithms are often defined in terms of two projection
operators $P_F^{\bm a}$ and $P_Q$ that project onto the solution $z$ to Eqs.
(\ref{eq:data_F} and \ref{eq:overlap}) that is closest to the current
estimate described in Section \ref{sec:standard}. 

Alternative approaches include formulating the problem 
as:
\begin{eqnarray}
\min_\psi \| \ba-|{\bm F} \Q \psi| \|
\label{eq:nonlin}
\end{eqnarray}
and solve it by standard unconstrained minimization algorithms such as
 conjugate gradient, Newton and quasi-Newton methods
\cite{ fienup,lbnlptyco,Thibault-Guizar}.  Efficient projection 
operators can be used if we formulate Eq. \ref{eq:nonlin} using the
frames $\bz$ as slack variables 
as discussed in Section \ref{sec:standard} and solve:
\begin{eqnarray}
\min_{\substack{\bz} }\| \ba-|\bm F \bz|\| 
\label{eq:nonlinconst}
\end{eqnarray}
with the conditions that $\bz$ satisfies Eq. (\ref{eq:overlap}) using projected
gradient, Newton and quasi-Newton methods\cite{ lbnlptyco,lbnlptycho1}.  .

Another approach uses a
$n^2\times n^2$ phase-space described in terms of a ``Wigner Distribution'' function\cite{batesptycho,Chapman1996}.
 More recently \cite{phaselift,Ohlsson,candes11}  a convex relaxation
 of the quadratic problem is obtained by lifting to an $n^2 \times
 n^2$ space and minimizing the rank of the matrix 

\subsection*{Main results}

The main contribution of this paper is the introduction of an
additional optimization step in the  measurement space, which with a
  dimensionality of $km^2$, is larger than the object space, $n^2$. It
  is aimed to deal with fluctuating intensities, position errors, poor
  calibration using multiplexing illumination, and an unknown offset
  (background) for every pixel but constant throughout the acquisition
  (or vice versa).  Specifically, instead of solving Eq.
(\ref{eq:nonlin}), we wish to minimize the gap between the measurement space
and the smaller  object space:
\begin{eqnarray}
\min_{\substack{\bz,|\bm F \bz|=\bm a}} \|[I-P_\Q] \bz \|,
\label{eq:gap}
\end{eqnarray}
where $I$ is the identity operator  and $P_Q $ 
represents a projection  onto the object space and will be described in the following section.

Recently it was proposed to use maxcut algorithms to solve a similar
problem \cite{Waldspurger}. Here we approach the problem differently. We start from the redundant measurement space and compute
pairwise comparison between frames before merging into the object space.

 In section \ref{sec:augmented} we consider  the case that the diffraction measurement $a_{(i)}$ is contaminated {(i.e. multiplied)} by a unknown scalar factor
 $\omega_{(i)}$.  We encounter this problem
 if the intensity or integration time of the incident
beam is unknown. If we fix the relative amplitude and phase within every frame $z_{(i)}$ 
and minimize the gap Eq. (\ref{eq:gap}) with respect to the vector $\bm \omega$ for a given $\bz$, we
can express Eq. (\ref{eq:gap}) as:
\begin{eqnarray}
\min_{\bm \omega} &&\bm \omega^\ast H \bm \omega
\\ \nonumber
H_{i,j}&=& z_{(i)}^\ast \left (
 \delta_{i,j} I-  \tfrac{Q_{(i)}Q_{(j)}^\ast}{e_i^T(\Q^\ast\Q)e_j}
\right )z_{(j)},
\end{eqnarray}
where the $k\times k$ matrix $H$ is calculated by computing the pairwise dot product between
overlapping frames. While this problem arises from  the need to account for intensity
fluctuations, it turns out to be a useful technique to improve the convergence
rate for large scale problems. The phase vector obtained by
normalizing $\bm\omega$ enables us to adjust for the relative phase
between frames and accelerate the rate of convergence
in iterative algorithms. 

We use a similar approach to optimize
perturbations of the illumination matrix $\Q$:  the position among
frames (Section \ref{sec:positions}). Alternative approaches optimizing  the positions
from the reduced object space have been proposed by others
\cite{Maiden:2012, fienup}. 
By minimizing the gap between measurement space and constraint, we obtain a first order correction
formula  that relies on pairwise scalar products between
neighboring frames. We expect a method based on pairwise
comparsions to work well in large scale problems  when long range
position drifts may arise. 

In Section \ref{sec:tests} we report the following numerical results:
\begin{itemize}
\item Exact reconstruction with intensity fluctuation given by the
  coefficients $\omega_{(i)}$ (see Fig. \ref{fig:fixI0}).
\item Accelerated convergence (Fig. \ref{fig:scaling}) even when no
  intensity fluctuation is present in the data (Table \ref{tab:tests})
\item Exact reconstruction with multiplexing using 4 simultaneous
  illuminations adding incoherently on the detector, with perturbation
  of the amplitudes (Fig. \ref{fig:multiplexing})
\item Position recovery (Figs. \ref{fig:positions}) of the illuminating probe.
\item Joint reconstruction of the sample and fluctuating background noise independent
  from the sample (Figs.  \ref{fig:bkgdata},\ref{fig:bkg}).
\item Exact reconstruction with missing (corrupted) data entries (see Fig. 
\ref{fig:bstop},\ref{fig:bstop1}).
\end{itemize}
In the following section \ref{sec:standard} we will describe the
standard operators commonly used in the literature.  

\section{Standard Projection algorithms \label{sec:standard}}
The projection operator $P^{\ba}_F$ mentioned in the previous
section is often known as the \textit{Fourier magnitude} projection
operator.  Applying this operator to a vector $\bz$ yields
\begin{eqnarray}
\label{projection_f}
P_F^{\ba} \bz = {\bm F}^{\ast} \left(\frac{{\bm F} \bz}{|{\bm F} \bz|}\cdot \ba\right) . \ \
\end{eqnarray}
where division and multiplication are intended as element-wise operations.
It is easy to verify that $|P^{\ba}_F \bz| = |\ba|$ and therefore $P^{\ba}_F \bz$
satisfies Eq. (\ref{eq:data_F}) for any $\bz$. We mention that
$P^{\ba}_F$ is a projection in the sense that
\begin{eqnarray}
P_F^{\ba} \bz = \arg \min_{\bar \bz} \|\bz_{(i)}-\bar \bz_{(i)}\|, 
\label{fixp3}
\\
\nonumber
\text{ subject to } | {\bm F} \bar \bz|=\ba,\, 
\end{eqnarray}
where $\|\,\|$ denotes the Euclidean norm. 
The matrix $\Q$ defines an orthogonal projection operator 
$P_\Q $ that projects any vector in 
$\mathbb{C}^{km^2}$ onto the range of $\Q$:
\begin{equation}
\label{projections}
P_Q = \Q (\Q^{\ast}\Q)^{-1}\Q^{\ast},
\end{equation}
when $(\Q^{\ast}\Q)^{-1}$ exists.
An alternative formulation uses projection operators that apply
 on the Fourier frames $\bm \tz=\bm F \bz$:
\begin{eqnarray*}
\tilde P_\Q &=&\bm F P_\Q \bm F^\ast, \quad
\tilde P_F^{\bm a}=\bm F P_{F}^{\bm a} \bm F^\ast.
\end{eqnarray*}
Line search stratrategies to solve Eq. (\ref{fixp3}) can be
implemented more efficiently using this formulation\cite{lbnlptyco}.

 In the simple alternating projection algorithm, the
approximation to the solutions of (Eqs. (\ref{eq:data_F}) and (\ref{eq:overlap})) are
updated by:
\begin{eqnarray}
\label{eq:ap}
  \bz^{(\ell)} &=& \left [ P_{\Q}P^{\ba}_F \right ] \bz^{(\ell-1)} ,
  %\\
%\label{eq:pseudoinv1}
%\psi^{(\ell)}&=&\arg \min_{\psi}  \| \bm z^{(\ell)}-\Q\psi\|.
\end{eqnarray}
 where typically the initial guess $\bz^{(0)}$ is a random vector. Clearly 
 $P_F^{\bm a}$ from (\ref{fixp3})  forces $\bz^{(\ell-1)}$ to have the right amplitude in the Fourier domain, and $P_{\Q}$ forces  $P_F^{\bm a}\bz^{(\ell-1)}$  to be located in the  range of $\Q$. We note that the projector $P_\Q$ can be expressed by computing  the running estimate of $\hpsi$
 denoted as $\psi^{(\ell)}$ 
\begin{eqnarray}
\label{eq:pseudoinv1}
\psi^{(\ell)}&=&\arg \min_{\psi}  \| P_F^{\ba}\bm z^{(\ell-1)}-\Q\psi\|,
\end{eqnarray}
which is solved  by taking Eq. (\ref{projections}) into account:
\begin{eqnarray}
\label{eq:trueinvreg}
 \psi^{(\ell)}&=& \left [(\Q^\ast \Q)^{-1} \Q ^\ast\right ]{P_F^{\ba}}\bz^{(\ell-1)},
\end{eqnarray}
when $(\Q^\ast\Q)^{-1}$ exists. Notice that $P_{\Q} P_F^{\ba}\bz^{(\ell-1)}=\Q\psi^{(\ell)}$. 

We mention two practical issues regarding the updating steps (\ref{eq:pseudoinv1})  and (\ref{fixp3}). First, since $(\Q^{\ast}\Q)^{-1}$ may not exist, one may introduce a regularization factor $\epsilon$ into (\ref{eq:pseudoinv1}) and update the
running estimate as:
\begin{eqnarray}
\label{eq:pseudoinvreg}
  \psi^{(\ell)} = (\Q^{\ast}\Q+\epsilon)^{-1}
\left ( \Q^{\ast}  P_F^{\ba}
  \bz^{(\ell-1)}+\epsilon \psi^{(\ell-1)} \right ).
\end{eqnarray}
with typically $\psi^{(0)}=0$.  $\epsilon$ is a
regularization factor that leaves unchanged the entries that are never
illuminated, and gradually reduces the correction from $\psi^{(\ell-1)}$ where the sum of the illuminating probe intensities is small. 
If we replace Eq. (\ref{eq:trueinvreg}) by Eq. (\ref{eq:pseudoinvreg}) we obtain an operator that is no longer a projection operator but can be viewed as a relaxed projection.

If the entries of $a_{(i)}(\bq)$ are corrupted by
  gaussian random noise  with known variance $\bm \sigma_{(i)}^2(\bq)$, which is expressed as a $m^2$ long column vector, one may
  replace Eq. (\ref{fixp3})
modifying $P_F^{\ba}$ with

\begin{eqnarray}
\label{eq:Pfas}
 P_F^{(\ba,\sigma)} \bz=\bm F^*
\left(  
\frac{\bm F \bz}{|\bm F \bz|}\cdot {\frac {\ba+|\bm F \bz| \bm\sigma^2} {1+\bm\sigma^2}}\right),
\end{eqnarray}
%where $\tilde \bz=\bm F \bz$. 
where $\bm \sigma$ is viewed as a
regularization factor that leaves unchanged the entries of $\bz$ when the data entries
are completely corrupted ($\bm \sigma^2\to \infty$). 
Clearly $\tfrac {\ba+|\bm F \bz|\bm\sigma^2} {1+\bm\sigma^2}\tfrac 1  {|\bm F \bz|} \to 1$ when $\bm \sigma^2\to \infty$. When $\bm \sigma^2\to 0$,   (Eq \ref{eq:Pfas}) reverts to the regular projection operator
(Eq. \ref{projection_f}).

%\sout{Second, if the entries of $a_{(i)}(\br)$ are corrupted by
%  gaussian random noise  with known variance $\bm \sigma_{(i)}^2(\br)$, which is expressed as a $m^2$ long column vector, one may
%  replace Eq. (\ref{eq:pseudoinv1}) by the weighted least square
%  version:}
%\begin{align} 
%\psi^{(\ell)}=& \arg \min_{\psi} \sum_{i}  \Diag{\tfrac 1 {1+\bm \sigma^2_{(i)}}} \left | {\cal
%    F} \left ( {\color{red} P_F^{\ba}}\bm z^{(\ell)}_{(i)}-
%    \Q _{(i)}\psi \right ) \right |^2
%\nonumber\\
%=&\sum_{i} \left (\sum_{\iota} {\cal F} \Diag{\tfrac 1 {1+\bm \sigma^2_{(\iota)}}} {\cal F}^* Q_{(\iota)}^{\ast}
%  Q_{(\iota)} \right )^{-1}\nonumber\\
%&\times Q_{(\iota)}^{\ast} {\cal F} \Diag{\tfrac 1 {1+\bm \sigma^2_{(\iota)}}} {\cal F}^*{\color{red} P_F^{\ba}} \bz^{(\ell)}.
%\label{eq:pseudoinvmask}
%\end{align}
%\sout{
%Clearly $\tfrac 1 {1+\bm\sigma^2_{(i)}}\to 0$ when $\bm \sigma^2_{(i)}\to \infty$, that is, when the data entries $\bm a$ are completely corrupted, we simply ignore that information.
%%where $\sigma_{(i)}$ is an $m^2$ long column vector. 
%The term in $()^{-1}$ can be precomputed. 
%}

The simple alternating projection algorithm can be viewed as the
projected steepest descent algorithms.  Projected conjugate gradient
methods have also been proposed in
\cite{fienup,lbnlptyco,Thibault-Guizar} to accelerate convergence rate.

 A number of heuristic first order algorithms have been proposed that outperform
 the simple algorithms,
a few examples are given in Tab. \ref{tab:hioraar}, with $\beta\in [0,
1]$ is a relaxation parameter. Very recently, an alternating direction
method (ADM) {was proposed} to work with a special augmented Lagrangian
function \cite{ZWen}. This function is minimized by applying a block coordinate
descent scheme\cite{zwen2} (or alternating search directions) akin to these
projection operators.

\begin{table*}[htbp]
  \center
  \begin{tabular}{|l|l|}
    \hline
    projection algorithm & updating formula $\bz^{(\ell+1)} =$ 
\\ 
\hline \hline
    Alternating Projection \cite{hio} & 
$\left [ P_QP_F \right ] \bz^{(\ell)}$
\\ 
\hline 
    HIO \cite{hio} & 
$\left [ P_QP_F +
    (I-P_Q)(I- \beta P_F)\right ] \bz^{(\ell)}$
\\ 
\hline 
    Difference Map \cite{Thibault2008} &  
$\left [ P_FP_Q +
    (I-P_F)(I- \beta P_Q)\right ] \bz^{(\ell)}$
 \\ \hline 
    RAAR\cite{raar} &
$  \left [ 2 \beta P_Q P_F+(1-2 \beta) P_F +\beta (P_Q-I)\right ]
  \bz^{(\ell)}$
\\ \hline
  \end{tabular}
\caption{
Popular fix-point algorithms used in phase retrieval. HIO: hybrid input-output algorithm. RAAR: Relaxed averaged alternating reflections algorithm.
\label{tab:hioraar}
}
\end{table*}

\section{Fluctuating intensities, 
and augmented phase retrieval\label{sec:augmented}}
Intensity fluctuations can be accounted for by introducing a scalar scaling
factor $\omega_i\in \mathbb{C}$ multiplying every ($m^2$) pixel of a
diffraction frame. This can be expressed in various forms (frame by
frame or all at once) as:
\begin{eqnarray}
|({\cal F} z_{(i)}) | \cdot |\omega_{(i)}|&=&a_{(i)},  \quad \forall i \\
| {\Diag{\bm F \bz} } \bm B\bm \omega|&=&\ba  ,\quad  (\bm B)_{i,j}=\delta_{i,j}  1_{m^2},
\label{eq:bins}
\end{eqnarray}
where $\bm\omega=[\omega_1,\ldots,\omega_k]\in\mathbb{C}^k$ and $\bm B$ is {a $k\times k$ diagonal block matrix with the diagonal entry $1_{m^2}$, which is the $m^2\times 1$ matrix with $1$ in all entries. In other words, $\bm B$} copies the scalar factors
$\omega_{i}$ before multiplying by $\bm F \bz$. 
In practice $\omega_i$ is unknown and needs to be estimated. If we know $\bz$ or its approximation, to estimate $\omega_i$, we find the vector $\bm \omega$ { that minimizes the gap}
with the  {object} space:
\begin{eqnarray}
\arg \min_{\bm \omega} &&
 \left \| (I-P_\Q) \Diag{\bz} \bm B \omega\right \|^2.
\label{minc2}
\end{eqnarray}
 We can write Eq. (\ref{minc2}) as:
\begin{eqnarray}
\label{eq:synchframes}
&&\arg \min_{\bm \omega} 
\bm \omega^\ast H \bm \omega,
\\ \nonumber
H_{i,j}&=& \bm 1^T \Diag{z_{(i)}^\ast} \left ( \delta_{i,j} I-  \frac{Q_{(i)}Q_{(j)}^\ast}{e_i^T(\Q^\ast\Q)e_j} \right )\Diag{z_{(j)}}\bm 1,
\\ \nonumber
&=& {z_{(i)}^\ast} \left ( \delta_{i,j} I-  \frac{Q_{(i)}Q_{(j)}^\ast}{e_i^T(\Q^\ast\Q)e_j} \right )z_{(j)},
\end{eqnarray}
% \left(P_Q \right )_{(i,j)} &=&Q_{(i)}\tfrac 1 {\Q^\ast \Q}Q^\ast_{(j)},
where the  $k\times k$ matrix $H$ is computed by performing the scalar product between every
pair of overlapping frames. We can eliminate the trivial solution $\bm \omega=0$ by setting an
additional constraint such as $\sum \bm \omega$, or $\|\bm \omega\|$=constant.
A simple way to solve this problem (see appendix \ref{sec:i0}) is to start with
$\bm 1$ as our first guess for $\omega$ and solve:
\begin{equation}
\label{eq:poweriter}
H \bm \omega= \alpha  \bm 1,\quad 
\end{equation}
where  $\alpha$ is chosen to normalize the average flux $\|\bm
\omega\|/\|\bm 1\|$.

%{\color{red}
%Another possibility is to set $\|\bm \omega\|$=constant and compute the largest eigenvalue of the %matrix
%\begin{equation}
%\label{eq:eigs}
%A_{i,j}= 
%\tfrac{\bz_{(i)}^\ast}{\sqrt{\|\bz_{(i)}\|}}
  %\tfrac{Q_{(i)}Q_{(j)}^\ast}{e_i^T(\Q^\ast\Q)e_j}
 %\tfrac{\bz_{(j)}}{\sqrt{\|\bz_{(j)}\|}\|}
%\end{equation}
%}

In order to take the intensity fluctuation problem into account while applying  the projection algorithm (\ref{eq:data_F}) and (\ref{eq:overlap}), we introduce the following solution.
First, we replace the operator $P_\Q$ used in the standard projection algorithms 
listed in Table~\ref{tab:hioraar} by an augmented projection operator
$P_\Q^{\bm \omega}$ defined as:
\begin{equation}
\label{eq:augproj}
P_{\Q}^{\bm \omega}= D^{-1}_{\bm \omega}  P_\Q D_{\bm \omega}, \quad
D_{\bm \omega}=\Diag{\bm \omega},
\end{equation}
where $\bm \omega$ is the solution to Eq.(\ref{eq:synchframes}) or
Eq.(\ref{eq:poweriter}). 
An alternative modification is to  recompute the normalization factor $\bm Q^\ast \bm Q$ with the scaling factors $|\omega_i|^2$. This yields an orthogonal projection: 
\begin{equation}
\bar P_{\Q}^{\bm \omega}= \Q_{\bm \omega} \left (\Q^\ast_{\bm \omega} \Q_{\bm \omega}
\right )^{-1} \Q_{\bm \omega}^\ast , \quad \Q_{\bm \omega}=D_{\bm \omega}^{-1}\Q
\end{equation}

When no intensity fluctuations are present,
we normalize $\bm \omega$ and replace $D_{\bm\omega}$ with:
$$
D_{\bm \hat \omega}=\Diag{\frac {\bm \omega} {|\bm \omega|}}.
$$

Although the construction of $P_{\Q}^{\bm \omega}$ is motivated by the need
to account for intensity fluctuations among different diffraction
frames in the measured data, it turns out to be a useful technique for
accelerating the convergence of projection algorithms even when no
intensity fluctuation is present in the data.  The minimization
problem in  (Eq. \ref{minc2}) is similar to the phase
problem of how to merge frames $\bz_{(i)}$ with unknown phase factor
$\bm \omega$, which can be written as:
\begin{equation}
\arg \min_{\bm \omega, |\omega_i|=1}  
  \bm \omega H \bm \omega. 
\label{minc1}
\end{equation}
Replacing the condition $|\bm \omega|=\bm 1 $
with weaker conditions such as $\sum \bm \omega$ or $\|\bm
\omega\|=$constant enables us to solve this problem more efficiently. 
A similar approach is discussed in \cite{singer}.

The problem of the incoherent superposition of different signals can be treated 
in a similar way. 

\subsection{\label{sec:multiplexing} Multiplexing and incoherent measurements}

The incoherent measurement model is as follows. We consider $z_i$ the
highly redundant set of frames generated for all the positions of the
illumination function during an exposure.  For example, a single
exposure $a_{(i)}$ may
represents the sum of the intensities generated by an illumination
beam that translates during the exposure, or may represent 
a binned sample of a continuous signal. {Assume we have $k$ redundant measurement $\bm z_{(i)}$, where $i=1,\ldots,k$. The incoherent measurement is introduced by summing $s<k$ illumination windows according to a weight factor, where we assume $k/s$ is an integer for convenience. The weight factor, or the integration
time, for each frame, is represented by $|\bm \omega|^2$. In particular, the redundant set of frames $z_{(i)}$ is not measured directly; instead it is multiplied by a known averaging operator, which is expressed as $|\bm \omega|^2$ and a $(k/s)m^2\times km^2$ real matrix $\bm\Omega$ with all non-zero entries $1$.  Geometrically, $\bm\Omega$ groups the frames, which are weighted by $ |\bm \omega|^2$, and then the weighted frames in
each group are summed.} The incoherent measurement can thus be expressed by modifying (\ref{eq:bins}):
\begin{eqnarray}
%\bm a^2&=&\bm \Omega \left | {\bm F} \bm z  \right |^2 \bm B \bm |\bm \omega|^2,
\bm a^2&=&\bm \Omega \left ( | \bm F \bz|^2 \cdot \bm B\bm |\omega|^2 \right ),
\label{eq:incoherent}
\end{eqnarray}
where $\cdot$ and $|\cdot|^2$ are
intended as elementwise operations $\bm z$ is a given $km^2$ complex column vector and $\bm a^2$ is a $(k/s)m^2$ real column vector.
The projection operator associated with this problem can be expressed as
follows:
\begin{align}
P_F^{a,\bm\Omega}  \bz= \bm F^\ast \left (\Diag{\sqrt{ \bm \Omega^\ast \frac{\bm 
a^2} {\bm \Omega \left (| \bm F \bz|^2 \cdot \bm B\bm |\omega|^2\right ) }}}\bm F \bz \right ),\label{incoherence:projection:operator}
\end{align}
when all non-zero entries of $\bm \Omega$ are $1$ and $\bm\Omega\bm\Omega^\ast=s\bm I_{(k/s)m^2}$.
Here $\bm \Omega^\ast$ copies the entries over all
the frames that contribute to an exposure $a_{(i)}$. We can directly check that Eq. (\ref{incoherence:projection:operator}) satisfies Eq. (\ref{eq:incoherent}). Replacing this operator in the
reconstruction process is a subject of recent interest by several groups
\cite{clark:2011, thibault:13}. Other approaches for incoherent averaging over wavelengths, orientation, coherence, etc.  have been discussed by others \cite{Fienup:93}, \cite{Abbey2008},  \cite{CDIpartialcoherence},\cite{spence-flip}.

If $|\bm \omega|^2$ is unknown, we can derive it from solving a minimization
problem of the type Eq. (\ref{eq:nonlin}), from the object space, but with an incoherent
measurement model. Another approach is to obtain $\bm\omega$ 
by solving a minimization problem of the type Eq. (\ref{eq:gap}), from
the measurement space with
the incoherent measurement model  by solving Eq.\ref{eq:poweriter} and using
Eq. \ref{eq:augproj} .

Numerical tests described in section \ref{sec:tests} show the exact
recovery (within numerical precision) of the object and a multiplexing
array of beam positions averaged incoherently with errors in the
calibration of the amplitude factors. The ability to recover the
relative amplitude of a redundant set of frames that are averaged
during the measurement enables us also to identify which instance of
the experimental parameters occurred, or to
recover or calibrate the amplitude coefficients of a  multiplexing
array of incident beams.   

The number of frames used in the calculation however increases, and with it, the computational
cost increases as well. To reduce the number of parameters to optimize we can describe the change
in measurement space using Taylor expansion.

\section{\label{sec:positions} Position retrieval}
We consider the case in which the probe $w$ is translated from the 
input coordinate by an unknown distance $\xi$. 
We call $\Q_\xi$ the unknown illumination matrix used to generate the data.
To determine the illumination matrix, we determine the parameter $\xi$
so that the error $\varepsilon_{\Q_\xi}$ is minimized:
\begin{eqnarray}
\label{eq:minpos}
%\arg\min_\xi &
\varepsilon_{\Q_\xi}:=\left \|
\left [I-P_{\Q_\xi }\right] \bz
\right \|^2.
\end{eqnarray}
Given the illumination function $w$, we can compute the first and
  second order derivatives with respect to translation. 
  
We denote by $Q_{(i)},R_{1,(i)}, R_{2,(i)}, S_{11,(i)},S_{12,(i)}, S_{21,(i)}$ and $S_{22,(i)}$ 
  the illumination matrices that extract a frame out of an image and
  multiply by $w_{(i)}(\br)$, $\partial_{x_1} w_{(i)}(\br)$,
$\partial_{x_2} w_{(i)}(\br), \partial_{x_1}^2 w_{(i)}(\br), \partial_{x_1,x_2}^2 w_{(i)}(\br), \partial_{x_2,x_1}^2 w_{(i)}(\br)$ and $\partial_{x_2}^2 w_{(i)}(\br)$ respectively.
Build up $\Q,\R_1,\ldots,\DDQ_{22}$ from $Q_{(i)},R_{1,(i)}, \ldots, S_{22,(i)}$, which are tall and skinny matrices of the same size
  as $\Q$ discussed earlier, with identical location of the non-zero entries.
Assume that $\Q_\xi$ satisfies the following second order perturbation from $\Q$:
\begin{align}
&\Q_\xi=\Q+\text{diag}(\bm B\bm\xi_1)\R_1+\text{diag}(\bm B\bm\xi_2)\R_2  \nonumber\\
&+\text{Diag}(\bm B{\bm\xi}^2_1)\DDQ_{11} +  2\text{diag}(\bm B(\bm\xi_1\cdot\bm\xi_2))\DDQ_{\times}
  +\text{Diag}(\bm B{\bm\xi}^2_2)\DDQ_{22}\nonumber
\end{align}
where $\cdot$ and $\cdot^2$ are
intended as elementwise operations, ${\bm\xi}_1$ (resp. ${\bm\xi}_2$) is a $k\times 1$ matrix so that the $i$-th entry is the translation distance in the $x$-axis (resp. $y$-axis) of the $i$-th illumination window,
and $\DDQ_{\times}\equiv\tfrac 1 2 \left (\DDQ_{12}+ \DDQ_{21} \right )$.
Using this Taylor expansion into Eq. (\ref{eq:minpos}) and setting $\partial_{\xi_1}^\ast \|\cdot \|$ and $\partial_{\xi_2}^\ast \|\cdot \|$ to 0 gives (see appendix \ref{sec:taylor} for the detailed derivation for the $1$-dim case.):
\begin{eqnarray}
\label{eq:positions}
\left (
\begin{array} {ll}
H_1 & H_{\times}\\
H_{\times}   & H_2
\end{array}
\right )
\left (
\begin{array} {c}
\xi_1\\
\xi_2
\end{array}
\right )=
\left (
\begin{array}{c}
\buz_{i}^\ast \bz_{\R_{1i}}+
 \bz_{\R_{1i}}^\ast \buz_{i}
\\
\buz_{i}^\ast \bz_{\R_{2i}}+
 \bz_{\R_{2i}}^\ast \buz_{i}
\end{array}
\right ),
\end{eqnarray}
using the definition $\bz_{\R_1,\dots,\DDQ_{22}}\equiv[\R_1,\dots,\DDQ_{22}] \tfrac 1 {\Q^\ast \Q}\Q^\ast \bz$, $\buz\equiv[I-P_\Q ] \bz$, and 
where the matrices $H_{1}$, $H_2$ and $H_{\times}$ are defined as:
\begin{eqnarray}
\nonumber
\left ( 
H_1
\right )_{ij}
&=&
  \left (
    \bz_{\R_{1i}}^\ast  \bz_{\R_{1i}}
    -2\buz_i^\ast   \bz_{\DDQ_{11i}} %    -\bz_{\DDQ_{11i}}^\ast   \buz_i 
  \right )
  \delta_{ij}
-\bz_i^\ast \left ( O_{11} \right)_{ij}
 \bz_j+\mathrm{cc}
,
\\
\nonumber
\left ( 
H_2
\right )_{ij}
&=&
  \left (
    \bz_{\R_{2i}}^\ast  \bz_{\R_{2i}}
    -2\buz_{i}^\ast   \bz_{\DDQ_{22i}} %    -\bz_{\DDQ_{22i}}^\ast\buz_i 
  \right )
  \delta_{ij}
-\bz_i^\ast \left ( O_{22} \right)_{ij} \bz_j+\mathrm{cc}
,
\\
\nonumber
\left (H_{\times} \right )_{ij}&=&\left (
\bz_{\R_{1i}}^\ast \bz_{\R_{2i}}
-2\buz_i^\ast \bz_{S_{\times_i}} %-\bz_{S_{\times_i}}^\ast \buz_i
 \right ) 
\delta_{ij}
-\bz_i^\ast \left ( O_{\times} \right)_{ij} \bz_j+\mathrm{cc}
,
\end{eqnarray}
where $\mathrm{cc}$ denotes the complex conjugate term,  
\begin{eqnarray}
\nonumber
 \left ( O_{11} \right)_{ij}\equiv
\left (\R_1\right )_i
 \tfrac 1 {\Q^\ast \Q}
\left( \R_1  \right )_j^\ast, \\
\nonumber
 \left ( O_{22} \right)_{ij}\equiv
\left (\R_2\right )_i
 \tfrac 1 {\Q^\ast \Q}
\left( \R_2  \right )_j^\ast, \\
\nonumber
 \left ( O_{\times} \right)_{ij}\equiv
\left (\R_1\right )_i
 \tfrac 1 {\Q^\ast \Q}
\left( \R_2  \right )_j^\ast.
\end{eqnarray}

The system of equations (Eq. (\ref{eq:positions}))  can be solved efficiently by sparse linear algebra
solvers.   The
entries of the equation are given by the dot product between frames
($\bz,\buz, \bz_{R_1},\dots, \bz_{S_{22}}$) with partial overlap and
scaling factors given by $\R_{1,2}\tfrac 1 {\Q^\ast \Q} {\R_{1,2}}^\ast$. 
 The terms $\buz^\ast \bz_{\DDQ}$ in $H$ are higher order corrections close to
the solution and can be neglected in practice. 
In Section \ref{sec:tests}
we will show that this method can recover the position perturbations
to numerical accuracy when the perturbations are { smaller} than the
probe width.

\section{Background noise\label{sec:background}}
For completeness we consider an unknown offset $b(\bm q)\geq0$ (background) added to each frame. {A similar problem is discussed in  \cite{thurman2009} where Thurman and Fienup consider  the case of a constant signal bias $b(\bq)=b(0)$. Here we extend this approach  to a fluctuating background that is different for every pixel but constant throughout the illumination window. When $b$ is constant, the method described here reverts to  \cite{thurman2009, guizar:bias}.}
We express the relationship between the frames $z_{(i)}$, the data $a_{(i)}$ and the background $b$ as:
\begin{equation}
\label{eq:bkg}
|\tz _{(i)}|^2+b=a_{(i)}^2, \,\,\,\tz_{(i)}={\cal F} z_{(i)}.
\end{equation}
A less trivial variation of the problem is when $b_{(i)}(\bq)\geq 0$ is different for every frame but the same for every pixel $\bq$.

At each iteration, we solve the following \textit {offset} minimization
problem with an additional scaling parameter:
\begin{eqnarray}
\nonumber
 \min_{b,\eta} \sum_i\left | \left |\tz_{(i)} \right |^2 - \left |\tz_{(i)}^{(\ell)} \right |^2 \right |^2\\ 
\text{ subject to $|\tz_{(i)}^{(\ell)}|^2=\eta^{(\ell)} \left ( a_{(i)}^2 - b^{(\ell)} \right ),$}
\label{eq:bkg_min}
\end{eqnarray}
where we set the initial value of $b^{(\ell=0)}=0$,  and  $\eta\in\mathbb{R}^{m^2}$ is a shrinkage parameter
that accounts for the fact that $|\tz^{(\ell)}|^2$
is on average smaller than $a^2$.  This is because $\tz^{(\ell)}$ is
obtained from a sequence of linear projections that reduce the overall
norm.  Since $\tz$ is smaller, the solution to the off-set projection
problem (\ref{eq:bkg_min}) is biased towards a larger offset.
Introducing the shrinkage parameter $\eta$ equal for every frame
provides the flexibility to avoid this problem.

  By solving for $\eta$ first,  we obtain the first and second order terms:
\begin{eqnarray}
\label{eq:pfb}
\eta^{(\ell)}&=&\frac{\sum_{i} d_{(i)}
\left|\tz_{(i)}^{(\ell)}\right |^2 }
{\sum_{i} d_{(i)}^2},
\end{eqnarray}
where $d_{(i)}=a_{(i)}^2-b$.
Solving for $b$ for a fixed $\eta$ gives
\begin{eqnarray}
\nonumber
b^{(\ell)}-b^{(\ell-1)}&=&\frac{1}{k} \sum_{i} \left (  d_{(i)}-
\left|\tz_{(i)}^{(\ell)}\right |^2 \frac 1 {\eta^{(\ell)}}\right ) \,
\\
\nonumber
&=& \left \langle  d_{(i)}\right \rangle -
\frac 1 {\eta^{(\ell)}}
 \left\langle \left | \tz_{(i)}^{(\ell)}\right |^2\right \rangle  \,
\\
\nonumber
&=& \left \langle  d_{(i)}\right \rangle -
\frac{\langle d_i^2 \rangle 
{\left \langle  \left | \tz_{(i)}^{(\ell)}\right |^2\right \rangle}
}
{\left \langle d_i \left | \tz_{(i)}^{(\ell)}\right |^2\right \rangle}.
\end{eqnarray}
To avoid strong perturbations, however, we set $\eta(\bq)=.8$ if $\eta(\bq)<0.8$.
When optimizing for a fluctuating offset ($b_{(i)}(\bq )=b_{(i)}(0)$ constant for
every frame), we simply replace the sum over $i$ with the sum over
$\bq$.  The update of $z$ is then computed as a regular Fourier
magnitude projection operator with an intensity offset:
$$
\tilde P_{ F}^{\left (a_{(i)}^2-b^{(\ell)} \right )} \tz^{(\ell)}_{(i)}=\tz^{(\ell)}_{(i)} 
\sqrt{ \tfrac{a_{(i)}^2-b^{(\ell)}} {\left |\tz^{(\ell)}_{(i)}\right |^2}},
$$
where we used the notation $\tilde P=  {\cal F} P {\cal F}^\ast$.

In the following section we will show that common projection methods
can recover the background even if the SNR is much smaller than 1.

\section{Numerical tests\label{sec:tests}} 
The object used to simulate the diffraction pattern is obtained from
an SEM image of a cluster of commercial 50 nm colloidal gold spheres. {The image is shown in Fig. \ref{fig:data_ball}.}
  The gray scale values were converted to a
sample thickness varying between 0 and 50 nm, and we assigned the
complex index of refraction of a  750 eV x ray photon going through 
an organic compound (PMMA).  Here the numerical
tests are done assuming periodic boundary conditions. These boundary
conditions ensure that every region of the object $\psi$
is illuminated with an equal number of overlapping frames, in other
words the null space of $Q$ is empty. We use frame width $16\times16$,
probe width $8$, step size $5$, number of frames $8\times8 \ldots 64\times64$, RAAR
algorithm, $\beta=.75$. {The initial guess of the phase chosen to be random. }
There is no padding of the illumination function shown in
Fig. \ref{fig:probe} (the intensity measurement is slightly under-sampled).

The  metrics $\varepsilon_F, \varepsilon_q$ used to monitor
progress are functions depending on $\bz^{(\ell)}$:
\begin{eqnarray*}
\varepsilon_F\left(\bz^{(\ell)}\right)&=&\tfrac {\left \| \left [ P_F - I\right ] \bz^{(\ell)} \right \|}{ \| \ba\| }
,\\
 \varepsilon_Q\left (\bz^{(\ell)}\right)&=&\tfrac {\left \| \left [ P_Q -I \right ]\bz^{(\ell)}  \right \|}{ \| \ba\|} 
\end{eqnarray*}
where $I$ is the identity operator. This has to be compared to $\varepsilon_0$, the error w.r.t the known solution:
\begin{eqnarray*}
 \varepsilon_0\left (\bz^{(\ell)}\right )&=& \tfrac 1 {\|\bm a\|} {\min_\varphi \left  \| e^{i\varphi} \bz^{(\ell)}-\Q \hpsi \right \|},
\end{eqnarray*}
where $\varphi$ is an arbitrary global phase factor.

We report the following observations

\begin{itemize}

\item \textbf{Fluctuating intensities: (Fig. \ref{fig:fixI0})} {The intensity fluctuation in this tests are $20\%$.}
By solving the new LSQ problem {introduced in (\ref{eq:augproj}),} we obtain accelerated convergence and
exact reconstructions every time we tested the problem, see (Fig.
\ref{fig:fixI0}). No degradation (above numerical precision)
introduced by intensities perturbed by $20\%$. 

\item \textbf{Scaling: (Fig. \ref{fig:scaling})}
{We show improved convergence in the larger scale problems. The results are summarized in Table \ref{tab:tests}.} As we increase the number of
frames, convergence slows down for standard projection operators.
{The parameters used in this simulation are $m=16$, $D_x=4$, $k$ varies and $n=kD_x+m$, where $D_x$ is the step size of the illumination windows.}

\item \textbf{Incoherent Multiplexing: (Fig. \ref{fig:scaling})}
Deconvolution of the incoherent sum of frames translated by 
3 times the illuminating beam width.

\item \textbf{Incoherent beams with fluctuations: (Fig. \ref{fig:multiplexing}).}
Deconvolution of the incoherent sum of frames translated by 
3 times the illuminating beam width, with unknown amplitude.

\item \textbf{Positions: (Fig. \ref{fig:positions})} Recovery of the positions
perturbed by an unknown factor randomly distributed between $\pm2.5$ pixels.

\item \textbf{Background: (Fig. \ref{fig:bkgdata}, Fig. \ref{fig:bkg})}
$\langle \|\bz_{(i)}\| \rangle_{i}/\|b\|=0.5$. In Figure
  \ref{fig:bkgdata},\ref{fig:bkg} we obtain exact reconstruction of
  the object and background (Background ratio $\|a\|/\|b\|=10^{-6}$).
  Exact recovery (within numerical precision) was obtained with
  step size $\delta x=3r$.  No degradation (above numerical precision) introduced by the background, nearly no influence on convergence rate. 

\item \textbf{Missing data: (Fig. \ref{fig:bstop}, Fig.
\ref{fig:bstop1})}
Exact recovery (within numerical precision) using 
Eq (\ref{eq:Pfas}). Frame size:32x32, number of frames: 16x16, step size:3.5 pixels.

\end{itemize}

\begin{table}[htbp]
  \center
  \begin{tabular}{|c|r|r|l|}
    \hline
    \# frames & clock time(s) & iteration & $\varepsilon_0^2$ \\ 
\hline 
standard &&& \\
\hline 
4$\times$4     &   0.7        &  121      &   $<$1e-11\\
8$\times$8     &   1.4         &  125      &   $<$1e-11\\
16$\times$16   &   4.9         &  144      &   $<$1e-11\\
24$\times$24   &  26.3        &  400      &   4.3e-10\\
32$\times$32   &  36.3        &  400      &   4.3e-4\\
48$\times$48   &   90.7       &  400      &   3.4e-4\\
64$\times$64   &   137.5       &  400      &   5.3e-3\\
\hline
augmented &&&\\
\hline
4$\times$4     &   1.9        &  138      &   $<$1e-11\\
8$\times$8     &   2.7         &  141      &   $<$1e-11\\
16$\times$16   &   6.5         &  138      &  $<$1e-11\\
24$\times$24   &   14         &  134      &  $<$1e-11\\
32$\times$32   &   25.6        &  139      &   $<$1e-11\\
48$\times$48   &   60.4       &  142      &   $<$1e-11\\
64$\times$64   &   96.2       &  149      &   $<$1e-11\\
\hline
  \end{tabular}
\caption{
Performance of projection algorithms using matlab R2012a 64-bit (maci64) (lapack version 3.3.1, MKL 10.3.5) on 2x2.2GHz Quad-core Intel xeon
using frames of dimension $16\times 16$
. \label{tab:tests}
}
\end{table}

\section*{Conclusions}
While  phase retrieval problems are notoriously difficult to solve numerically, 
the high redundancy in ptychographic data enables not only robust
phase recovery\cite{Thibault-Guizar,Godard:12} but the recovery of other parameters such as the
illuminating function itself\cite{thibault:probe}, position 
\cite{Maiden:2012, fienup,axel}, coherence function \cite{clark:2011}, etc.

In this paper we introduce a modified projection operator for the
ptychographic reconstruction problem that accounts for fluctuating
intensities, position errors, partial coherence or poor calibration using multiplexing
illumination,  and an unknown offset (background) {difference} for every
pixel but constant throughout the acquisition (or vice versa). Our
approach starts from the redundant measurement space and computes
pairwise comparison between frames before merging into the object space.
 We describe first order methods to minimize the gap between measurement space and object space w.r.t. the perturbation parameter. We show that our method enhances the convergence rate of common projection algorithms. We show several cases where missing information (phases, bad pixels, positions, incoherence,  etc.) was retrieved exactly (to within numerical precision) starting from random phases.   
 This method appears to be robust when the amount of overlap between neighboring frames is around 50\% or more. 
 
 Further theoretical analysis on the relative merits between object-space minimization and measurement space minimization will be the subject of the future work.

{Here some numerical details deserve further developments.} By optimizing the phase of each frame from the redundant measurement space, we solve
the phase problem at a resolution given by the step size between
frames. This intra-frame phase optimization may be applied to merge
subregions reconstructed independently by distributed computer
systems. For three dimensional objects, we could apply similar approach to
 merge two dimensional views
reconstructed independently into one three dimensional object.
Finally, this intra-frame optimization could be applied to multi-scale
reconstructions where frames are divided in regions of Fourier space, or it could be applied to correct low order phase aberrations between frames.  More work is needed to  to establish
the optimal frequency of communication and the amount of overlap between
sub-reconstruction regions.

\section*{Acknowledgments}
The authors thank Prof. Bin Yu and Jeff Donatelli for discussions. 
This research was supported in part by
the Applied Mathematical Sciences subprogram of the Office of
Energy Research, U.S. Department of Energy, under contract
DE-AC02-05CH11231 (SM,CY), and by the 
Laboratory Directed Research and Development Program of Lawrence
Berkeley National Laboratory under the U.S. Department of Energy
contract number DE-AC02-05CH11231 (A.  S.), and by the
Director, Office of Science, Advanced Scientific Computing Research,
of the U.S. Department of Energy under Contract No. DE-AC02-05CH11231
(F.M.).  The computational results presented were obtained at the
National Energy Research Scientific Computing Center (NERSC), which is
supported by the Director, Office of Advanced Scientific Computing
Research of the U.S. Department of Energy under contract number
DE-AC02-05CH11232. Hau-tieng Wu acknowledges the support by Purdue National Science Foundation of United States (CCF-0939370) and Focused Research Group (DMS-1160319).

\widetext
\appendix
\section{Intensity fluctuations\label{sec:i0}}

One of the practical issues one may face in ptychography is the
intensity fluctuation among different diffraction frames introduced by
instabilities in the light source, optics and shutters.  Such
fluctuation can be accounted for by introducing a scaling factor
$\omega_i\in \mathbb{C}$ for each diffraction frame,  
As a result, the definition $\hat{z}_i$ is modified so that the
equation
\begin{equation}
\omega_{i} z_{(i)} = Q_{(i)} \hpsi, \label{cqpsi}
\end{equation}
holds for $i=1,2,...,k$.  

Since both $\omega_{i}$ and $\hpsi$ are unknown 
in (\ref{cqpsi}), the solution to (\ref{cqpsi}) is clearly not unique. 
To exclude the trivial solution $\omega_i = 0$, for $i=1,2,...k$ and $\psi =0$,
we introduce an additional constraint and solve
\begin{eqnarray}\label{lsqr_c}
  \left ( \psi_{\min}, \omega_{(i)} \right )=\arg \min_{\psi, \omega_{(i)}} \sum_{i} \left \| Q_{(i)} \psi-\omega_{i}
  z_{(i)} \right \|^2\\ 
\nonumber\text{ subject to $\sum_{(i)} \omega_{i}=\sum_{(i)} 1=k$},
\end{eqnarray}
which is equivalent to solve 
\begin{equation}
\min_{\psi,\omega_{i},\lambda}{\cal L}(\psi,\omega_i,\lambda),\,\text{where }\,{\cal
  L}= \sum_{i} \|Q_{(i)} \psi - \omega_{i} z_{(i)}\|^2 +2 \lambda\left(\sum_{i} \omega_{i}
  -\|\bm 1\|^2 \right), 
\label{eq:lagrangian}
\end{equation}
where $\lambda$ is a Lagrange multiplier. To find the coefficients $\omega_{i}$, we use
the normal equation associated with the LSQ problem (Eq. \ref{eq:lagrangian}) :
\[
\left(
\begin{array}{cccccc}
\sum_{i} Q_{(i)}^\ast Q_{(i)} & -Q_1^\ast z_1     & \hdots & \hdots       &-Q_k^\ast z_k    & 0      \\
-z_1^{\ast}Q_1  & z_1^{\ast}z_1 & 0      &  \hdots       &0 & 1 \\
\vdots          &  0            & \ddots  &              &\vdots & \vdots\\
\vdots          &  \vdots       &  &  & 0 &  \\
-z_k^{\ast}Q_k  &  0        & \hdots& 0 &z_k^{\ast}z_k & 1 \\
0               & 1             & \hdots &   & 1& 0 
\end{array}
\right)
\left(
\begin{array}{c}
\psi \\
\omega_1  \\
\vdots \\
\omega_k \\
\lambda
\end{array}
\right)= 
\left(
\begin{array}{c}
0 \\
0  \\
\vdots \\
0 \\
\|\bm 1\|^2
\end{array}
\right).
\]
We can partition the equation above as
%%%%%%%%%%%%%%%%%%%%%%%%%%%%%%%
{
\[
\left(
\begin{array}{ccc}
A & B^\ast &\bm 0 \\
B & D & \bm 1\\
\bm 0 &  \bm 1^\ast & \bm 0
\end{array}
\right)
\left(
\begin{array}{c}
\psi \\
{\bm \omega}\\
\lambda
\end{array}
\right)
=
\left(
\begin{array}{c}
0\\
0\\
 \| \bm 1\|^2 \\
\end{array}
\right),
\, 
\]
where
\[
A = \sum_{i=1}^k Q_{(i)}^\ast Q_{(i)},\ \
B = \left(
\begin{array}{c}
-z_1^{\ast} Q_1 \\
\vdots \\
-z_k^{\ast} Q_k 
\end{array}
\right), 
\ \
D = \left(
\begin{array}{ccccc}
z_1^{\ast}z_1                 & 0      &  \hdots       &  0  \\
0                             & z_2^{\ast}z_2  & \ddots  &  \vdots   \\
\vdots                        & \ddots & \ddots  &  0   \\
0                             &   \hdots     &       0 & z_k^{\ast}z_k     \\
\end{array}
\right),
\quad
\bm \omega=
\left(
\begin{array}{c}
\omega_{1}\\
\vdots\\
\omega_{k} 
\end{array}
\right).
\]
By the block factorization
\begin{eqnarray}
\label{eq:solution_c}
\left(
\begin{array}{ccc}
I       & 0 &0\\
-BA^{-1} & I&0\\
0       & -\bm 1^\ast H^{-1}&1\\
\end{array}
\right)
\left(
\begin{array}{ccc}
A & B^\ast & \bm 0\\
\bm 0  & H &\bm 1\\
\bm 0  & \bm 0 &-\bm 1^\ast H^{-1} \bm 1\\
\end{array}
\right)
\left(
\begin{array}{c}
\psi \\
{\bm \omega}\\
\lambda
\end{array}
\right)
=
\left(
\begin{array}{c}
0\\
0\\
\|\bm 1\|^2 \\
\end{array}
\right),
\end{eqnarray}
where the Schur complement $H=D-BA^{-1}B^\ast $ is given by
\begin{equation}
\nonumber
H_{i,j}= z_{(i)}^\ast \left (
 \delta_{i,j} I- Q_{(i)}  \left(\sum_{(\iota)} Q_{(\iota)}^\ast Q_{(\iota)} \right )^{-1}  Q_{(j)}^\ast  \right ) z_{(j)}.
\end{equation}

By block-wise inversion of Eq. \ref{eq:solution_c}, we obtain $\bm \omega$ and the scaling factor $\lambda$ from a sparse linear equation:
$$
H \bm \omega=-\lambda  \bm 1 ,\qquad \lambda=-\tfrac { \|\bm 1 \|^2} {\bm 1^{\ast} H^{-1} \bm 1}\,.
$$

{
\subsection*{Eigenvalue method}
If we make a change of variable,  $\nu_{(i)}=\|z_{(i)}\| \omega_{(i)}$, we can re-write the problem
$\arg\min_{\bm \omega} \bm \omega^\ast H \bm \omega
$ as:
\begin{eqnarray}
\arg\min_{\bm \nu} &&\quad\|\bm \nu\|^2-\bm \nu^\ast K \bm \nu,  \qquad 
K_{i,j}= \frac {z_{(i)}^\ast }{\|z_{(i)}\|}Q_{(i)}  \left(\sum_{(\iota)} Q_{(\iota)}^\ast Q_{(\iota)} \right )^{-1}  
 Q_{(j)}^\ast \frac { z_{(j)} }{\|z_{(j)}\|}.
\nonumber
\end{eqnarray}
The solution to this problem assuming $\|\nu\|$=constant is the eigenvector corresponding to the largest eigenvalue of the sparse matrix $K$. This can be computed efficiently using packages such as\cite{arpack}
}
}

\section{Taylor expansion\label{sec:taylor}}
We consider the case in which the probe $w$ is translated from the 
input coordinate by an unknown distance $\xi$. {We restrict ourselves to the $1$-dim ptychography problem to simplify the discussion.}
We call $\Q_\xi$ the unknown illumination matrix used to generate the data.
To determine the illumination matrix, we determine the parameter $\xi$
so that the error $\varepsilon_{\Q_\xi}$ is minimized:
\begin{eqnarray}
\label{eq:min0}
\arg\min_{\xi\in\mathbb{R}^k} &
\left \|
\left [I-P_{\Q_\xi }\right] \bz
\right \|^2&,
\end{eqnarray}
{where the $i$-th entry of $\xi\in\mathbb{R}^k$ represents the translation distance
 of the $i$-th frame}. 
Given the illumination function $w$, we can compute the first and
  second order derivatives with respect to translation. 
 We denote by $Q_i,R_i,S_i$
  the illumination matrices that extract a frame out of an image and
  multiplies by ($w(\bx),\partial_x w(\bx)$, $\partial_x^2 w(\bx)$)
  respectively. {Build $\Q,\R,\DDQ$ from $Q_i,R_i,S_i$, which} are tall and skinny matrices of the same size
  as $\Q$ discussed earlier, with identical location of the non-zero entries.
{Assume that} the probe is perturbed to second order as follows:
$$
\Q_\xi=\Q+\xi\R +\xi^2\DDQ.
$$
where, by a slight abuse of notation, $\xi$ denotes a {diagonal and real matrix so that the $i$-th diagonal entry, denoted as $\xi_i$ represents the translation distance
 of the $i$-th frame. With $\Q_\xi$ plugged into (\ref{eq:min0}), we now} minimize
\begin{eqnarray}\label{eq:min}
\arg\min_{\xi\in\mathbb{R}^k}&
\small{
\left \|
\left [
I- \left ( 
\Q+\xi \R+\xsi \DDQ
\right )
\left [
\left ( 
\Q+\xi \R+\xsi \DDQ
\right )^\ast
\left ( 
\Q+\xi \R+\xsi \DDQ
\right )
\right]^{-1} 
\left ( 
\Q+\xi \R+\xsi \DDQ
\right )^\ast
\right ]
\bz
\right\|^2.
}&
\end{eqnarray}
By Taylor expansion:
$$
[\cdot]^{-1}\simeq \frac{1}{\Q^\ast \Q}\left( 1-(\R^\ast \xi \Q+\Q^\ast \xi \R+O(\xi^2))\frac{1}{\Q^\ast \Q}\right),
$$
{when $\Q^\ast \Q$ is invertible. The second order} term $O (\xi^2)$
includes other second order terms that we will not need. 
We write
the expansion of the residual in  Eq. (\ref{eq:min}) $f_0+f_1(\xi)+f_2(\xi^2)$ as:
\begin{eqnarray}
f_0&=&[I-P_\Q]\bz \equiv\buz 
\end{eqnarray}
We define $\phi^\ast\equiv \frac 1 {\Q^\ast \Q} \Q^\ast$, and express
the first order as
\begin{eqnarray}
\nonumber
f_1(\xi)&=&\left [-\xi \R \phi^\ast 
    -\phi  \R^\ast \xi
    +\phi (\R^\ast \xi \Q+\Q^\ast \xi \R)
    \phi^\ast 
\right ] \bz. 
\end{eqnarray}
By defining $\bz_\R\equiv \R \phi^\ast \bz$, using 
$P_\Q=\Q \phi^\ast=\phi \Q^\ast$, and  rearranging, we get 
\begin{eqnarray}
\nonumber
f_1(\xi)&=&  - \xi \bz_\R- \phi \R^\ast \xi (\bz -\Q \phi^\ast  \bz) 
+\phi \Q^\ast \xi\bz_\R \\
&=& -(1-P_\Q) \xi \bz_\R- \phi \R^\ast \xi \buz.
\label{eq:f1}
\end{eqnarray}
By using the equality $\buz (I-P_Q)=\buz$,  setting $ O_\R \equiv
\R \phi^\ast  \phi \R^\ast=
\R \frac 1 {\Q^\ast \Q} \R^\ast
$ and rearranging, we obtain:
\begin{eqnarray}
\nonumber
f_2(\xi)&=&-\left [ \xi \R \tfrac{1}{\Q^\ast \Q} \R^\ast \xi
    +\xsi \DDQ \phi+
\Q O \left( \xi^2 \right)
\right ] \bz\\
&=& - \xi O_\R \xi \bz
    - \xsi \bz_\DDQ+
\Q O \left( \xi^2 \right)\bz
\label{eq:f2}
\end{eqnarray}
where $\bz_\DDQ\equiv \DDQ \phi^\ast \bz$.
We rewrite  Eq. (\ref{eq:min}) above as:
$$
\left \|
\left [I-P_{Q_\xi }\right] \bz
\right \|^2=
f_0^\ast f_0+f_0^\ast f_1+f_1^\ast f_0+f_1^\ast f_1+f_0^\ast f_2+f_2^\ast f_0+O(\xi^3).
$$
Note that $\buz^\ast \Q=\buz^\ast \phi=0$. Set $\bz_\Q\equiv \Q \phi \bz$ and obtain the first and second order terms of Eq. (\ref{eq:min}):
\begin{eqnarray}
f_0^\ast f_1+f_1^\ast f_0&=&-\buz^\ast \xi \bz_\R,  
- \bz_\R^\ast \xi\buz
\\
\nonumber
f_1^\ast f_1+f_0^\ast f_2+f_2^\ast f_0&=&
  \bz^\ast_\R \xi (I-P_\Q) \xi \bz_\R
+\buz\xi O_\R  \xi \buz
-\buz^\ast \xi O_\R \xi \bz
-\buz^\ast \xsi \bz_\DDQ
-\bz^\ast \xi O_\R \xi \buz
-\bz_\DDQ^\ast \xsi \buz,
\\
\nonumber
&=&\bz^\ast_\R \xi^2 \bz_\R
-\buz^\ast \xsi \bz_\DDQ-\bz_\DDQ^\ast \xsi \buz
-\bz^\ast \xi O_\R \xi \bz
+\bz_{\Q}^\ast \xi O_\R  \xi \bz_{\Q}
-\bz^\ast_\R \xi P_\Q \xi \bz_\R,
\end{eqnarray}
By using the definition of $\bz_\Q$, $\bz_\R$, $P_\Q$ and $O_\R$ it is easy to show that $\bz_{\Q}^\ast \xi O_\R  \xi \bz_{\Q}
=\bz^\ast_\R \xi P_\Q \xi \bz_\R$ and simplify as:
\begin{eqnarray}
f_1^\ast f_1+f_0^\ast f_2+f_2^\ast f_0
&=&\bz^\ast_\R \xi^2 \bz_\R
-\buz^\ast \xsi \bz_\DDQ-\bz_\DDQ^\ast \xsi \buz
-\bz^\ast \xi O_\R \xi \bz.
%&\simeq&\bz^\ast_\R \xi^2 \bz_\R-\bz^\ast \xi O_\R \xi \bz.
\end{eqnarray}

By setting $\partial_{\xi_i} \|\cdot \|^2=0$ {in Eq. (\ref{eq:min})}, we obtain the
linear equation {for solving $\xi$}:
\begin{eqnarray}
\nonumber
\sum_j\left (2
    \left (
      \bz_{\R_i}^\ast  \bz_{\R_i} 
      -\buz_{i}^\ast   \bz_{\DDQ_i}
      -\bz_{\DDQ_i}^\ast \buz_{i}  
    \right )\delta_{ij}
    -\bz_i^\ast  O_{\R_{ij}}  \bz_j 
    -\bz_j^\ast  O_{\R_{ji}}  \bz_i
  \right )
  \xi_j
  &=&
  \buz_i^\ast \bz_{\R_i}+
  \bz_{\R_i}^\ast \buz_i
\end{eqnarray}
We note that $-\buz^\ast \xsi \bz_\DDQ-\bz_\DDQ^\ast \xsi \buz$ is a
second order correction if we assume that $\bz$ is in that range of
an unknown $\Q_\xi$ for small $\xi$. {We thus have the following approximation:}
\begin{eqnarray}
f_1^\ast f_1+f_0^\ast f_2+f_2^\ast f_0\simeq\bz^\ast_\R \xi^2 \bz_\R
-\bz^\ast \xi O_\R \xi \bz.
\end{eqnarray}
{We can thus consider solving the following approximation equation in practice:
\begin{eqnarray}
\nonumber
\sum_j\left (2
      \bz_{\R_i}^\ast  \bz_{\R_i} \delta_{ij}
    -\bz_i^\ast  O_{\R_{ij}}  \bz_j 
    -\bz_j^\ast  O_{\R_{ji}}  \bz_i
  \right )
  \xi_j
  &\simeq&
  \buz_i^\ast \bz_{\R_i}+
  \bz_{\R_i}^\ast \buz_i
\end{eqnarray}
}

Extension to the two dimensional case is given in Section \ref{sec:positions}.

\newpage

\begin{figure}
\begin{minipage}[p]{0.45\linewidth}
\centering
    \includegraphics[width=1\linewidth]{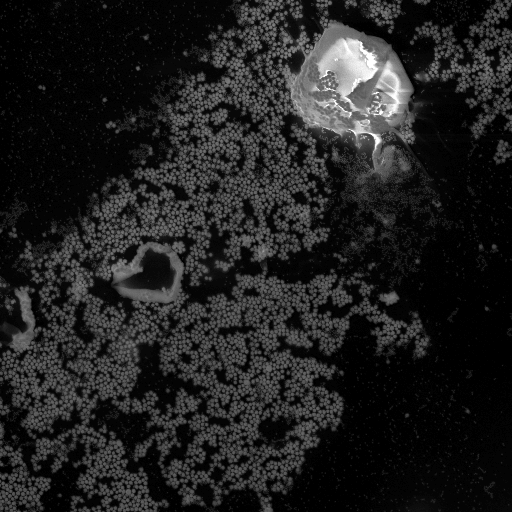}
\caption{object $\psi$ used to simulate diffraction data)}\label{fig:data_ball}
\end{minipage}
\hspace{0.5cm}
\begin{minipage}[p]{0.45\linewidth}
\centering
    \includegraphics[width=1\linewidth]{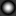}
\caption{Absolute value of the probe $\left |w(\bm r) \right |$ used in simulations (16$\times$16 pixels)}\label{fig:probe}
\end{minipage}
\end{figure}

\begin{figure}[hbtp]
  \begin{center}
    \includegraphics[width=0.4\textwidth]{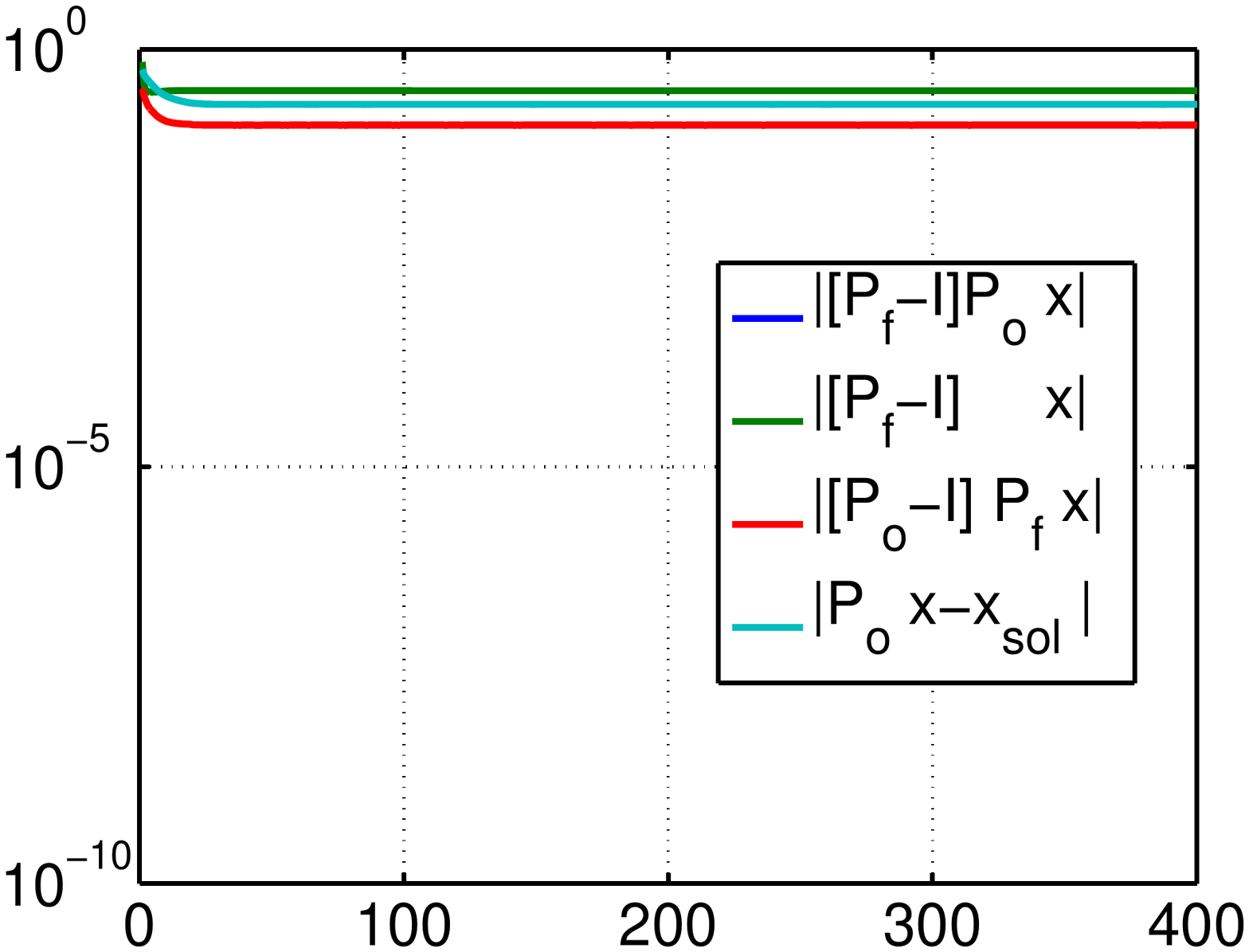}
    \includegraphics[width=0.4\textwidth]{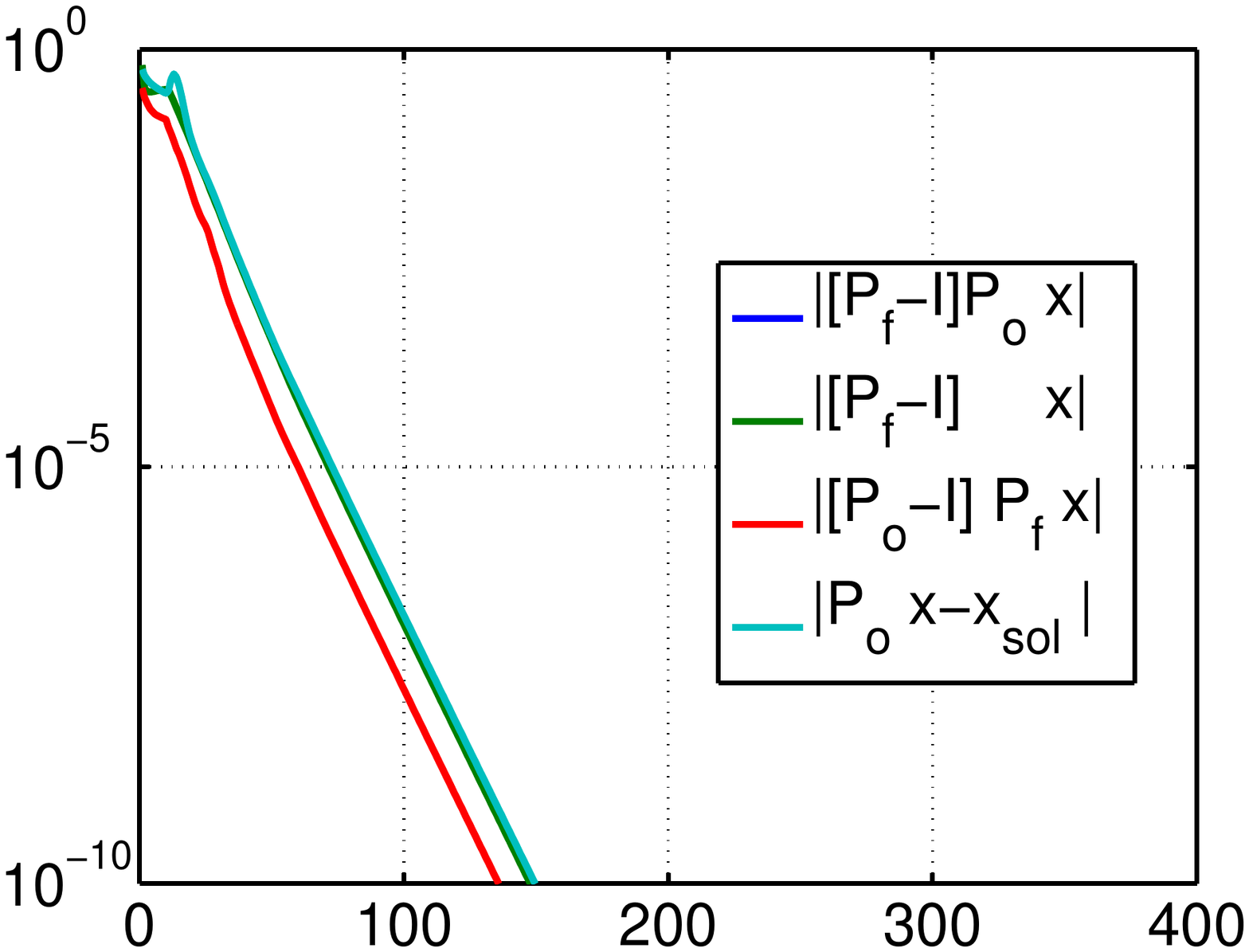}
    \includegraphics[width=0.25\textwidth]{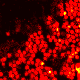} \ \ \ \ \ \ \ \ \ \ \ \ \ \ \ \ \ \ \ \ \ \ \ 
    \includegraphics[width=0.25\textwidth]{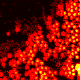}
 \end{center}
 \caption{ Convergence rate with an I0-error of $\pm20\%$. 
(left) old projection
   operator (right) new projection operator. (bottom) reconstruction
   from data with I0-error, and solution (reconstruction using the new
   projection operator is within the computer numerical precision, 
   { i.e. the figure on the right looks identical to the exact solution.  }
  \label{fig:fixI0}}
\end{figure}

\begin{figure}[hbtp]
\centering
\subfigure[8$\times$8 frames]{
    \includegraphics[width=.22\linewidth]{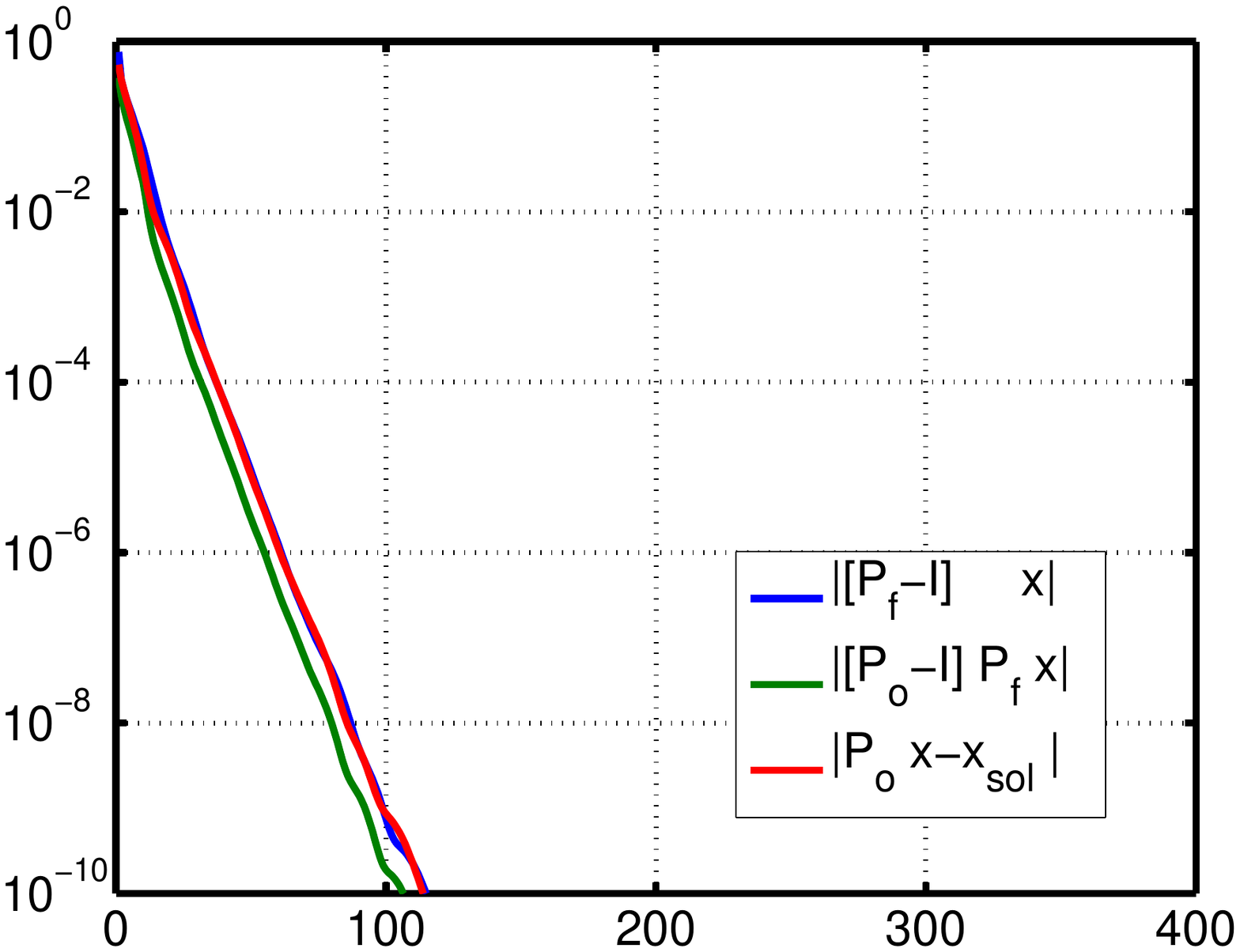}
}
\subfigure[16$\times$16 frames]{
    \includegraphics[width=.22\textwidth]{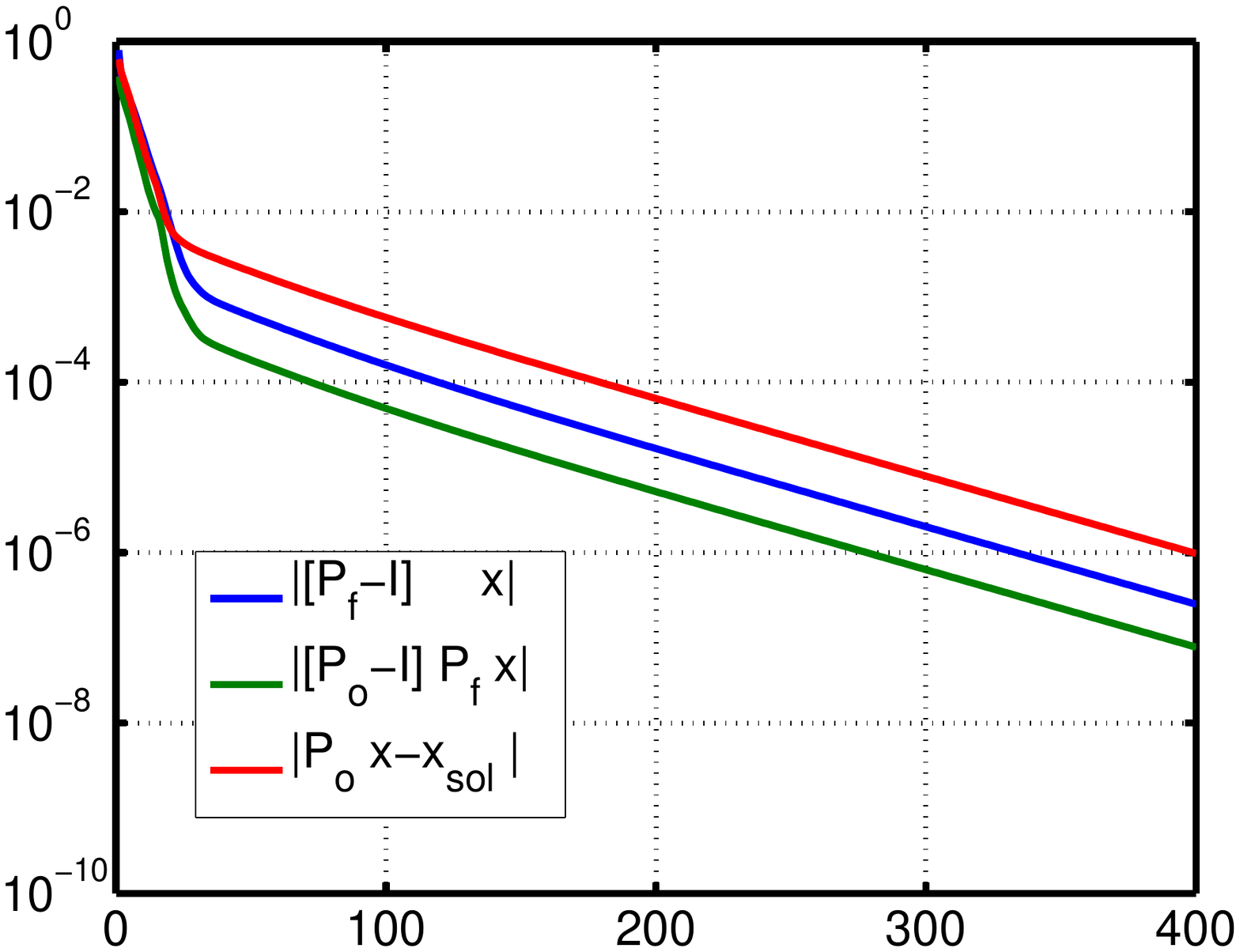}
}
\subfigure[32$\times$32 frames]{
    \includegraphics[width=.22\textwidth]{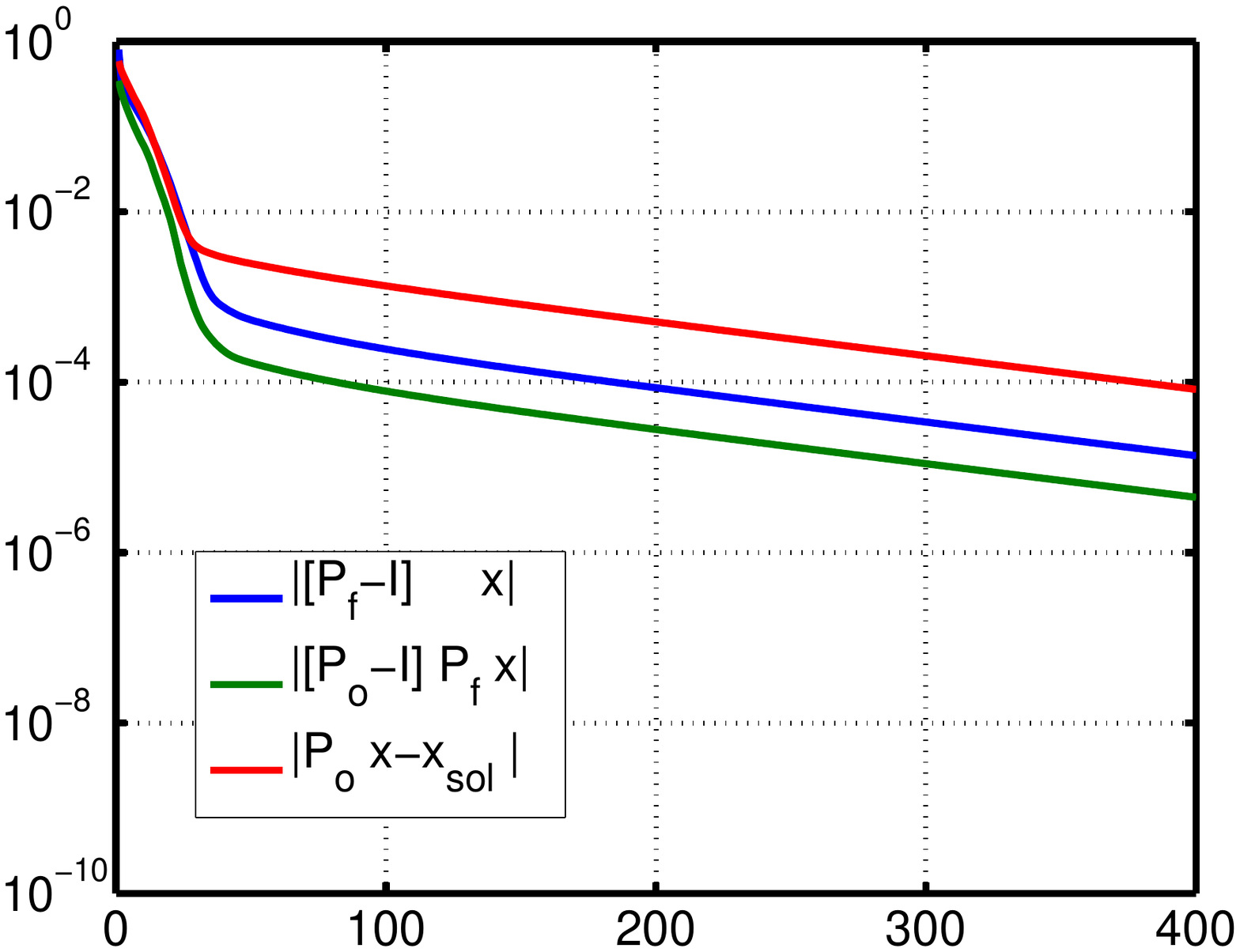}
}
\subfigure[64$\times$64 frames]{
    \includegraphics[width=.22\textwidth]{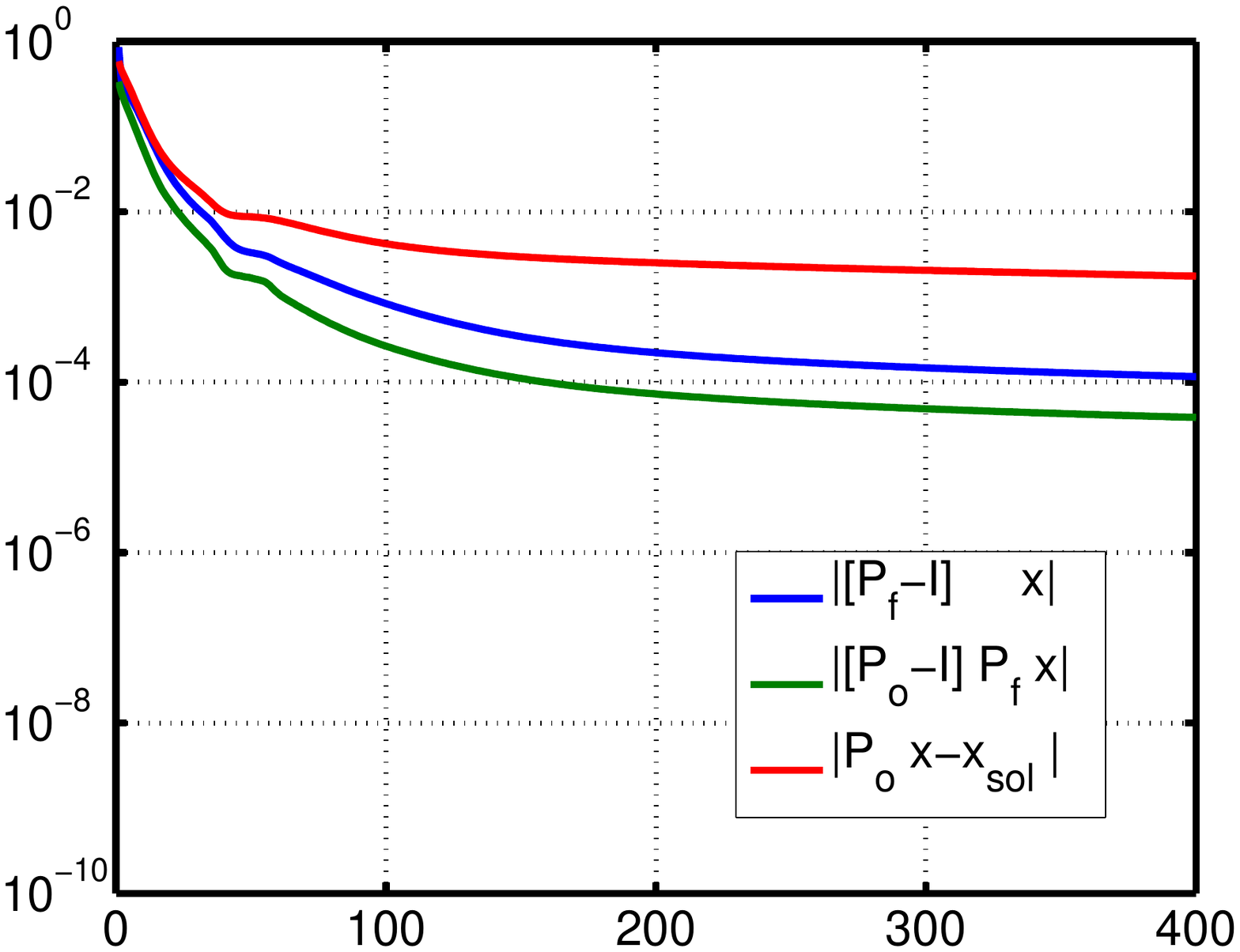}
}
\subfigure[8$\times$8 frames]{
    \includegraphics[width=.22\textwidth]{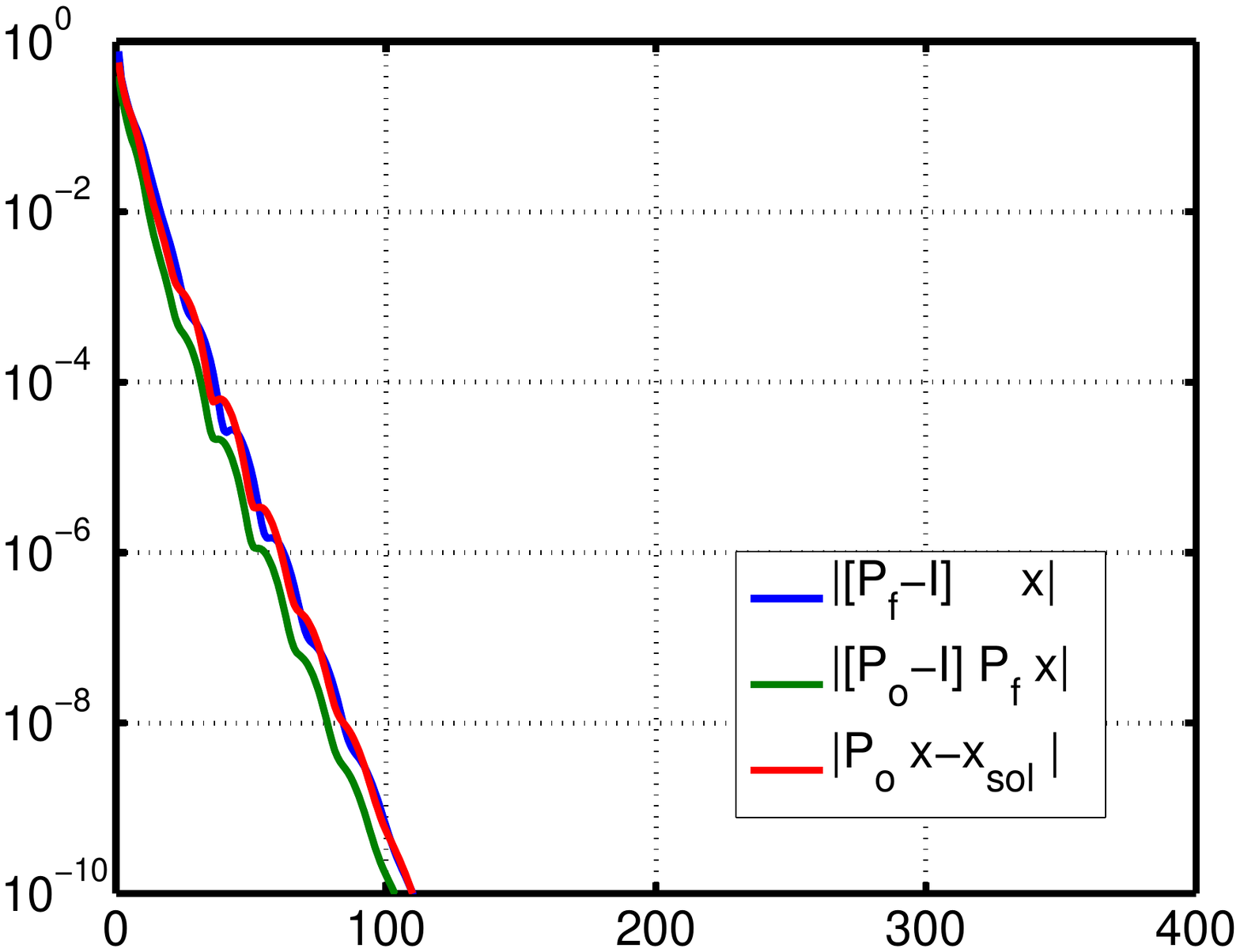}
}
\subfigure[16$\times$16 frames]{
    \includegraphics[width=.22\textwidth]{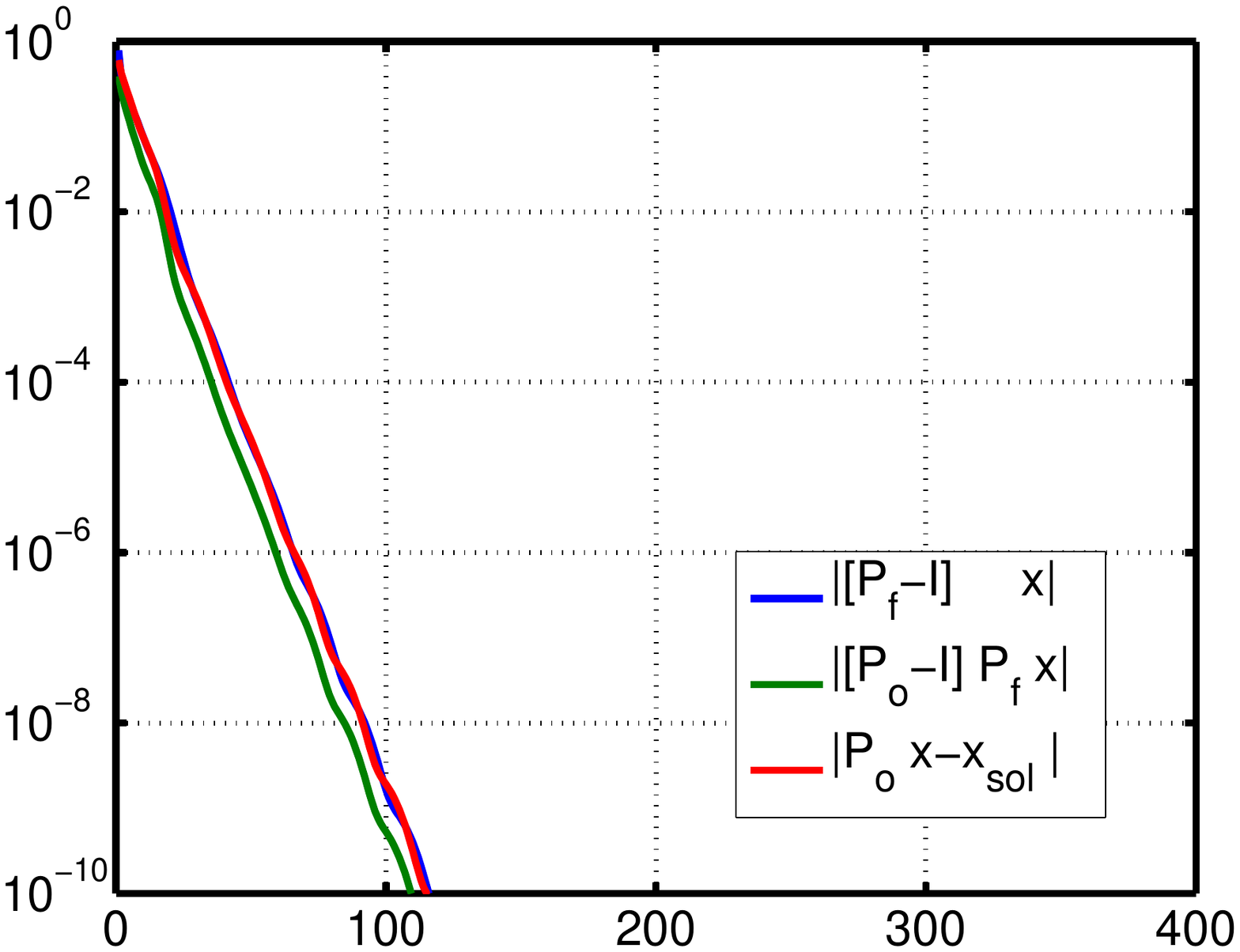}
}
\subfigure[32$\times$32 frames]{
   \includegraphics[width=.22\textwidth]{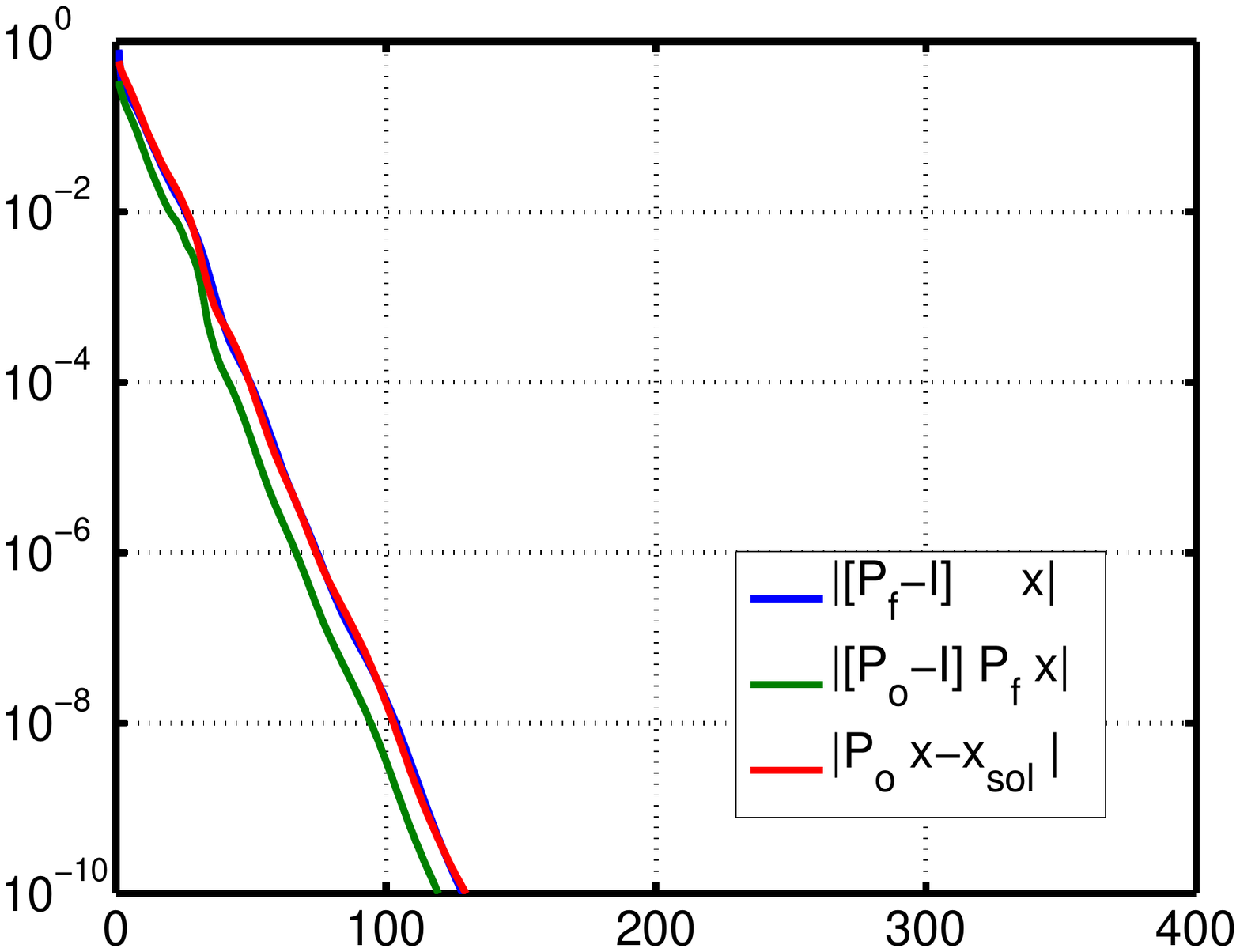}
}
\subfigure[64$\times$64 frames]{
    \includegraphics[width=.22\textwidth]{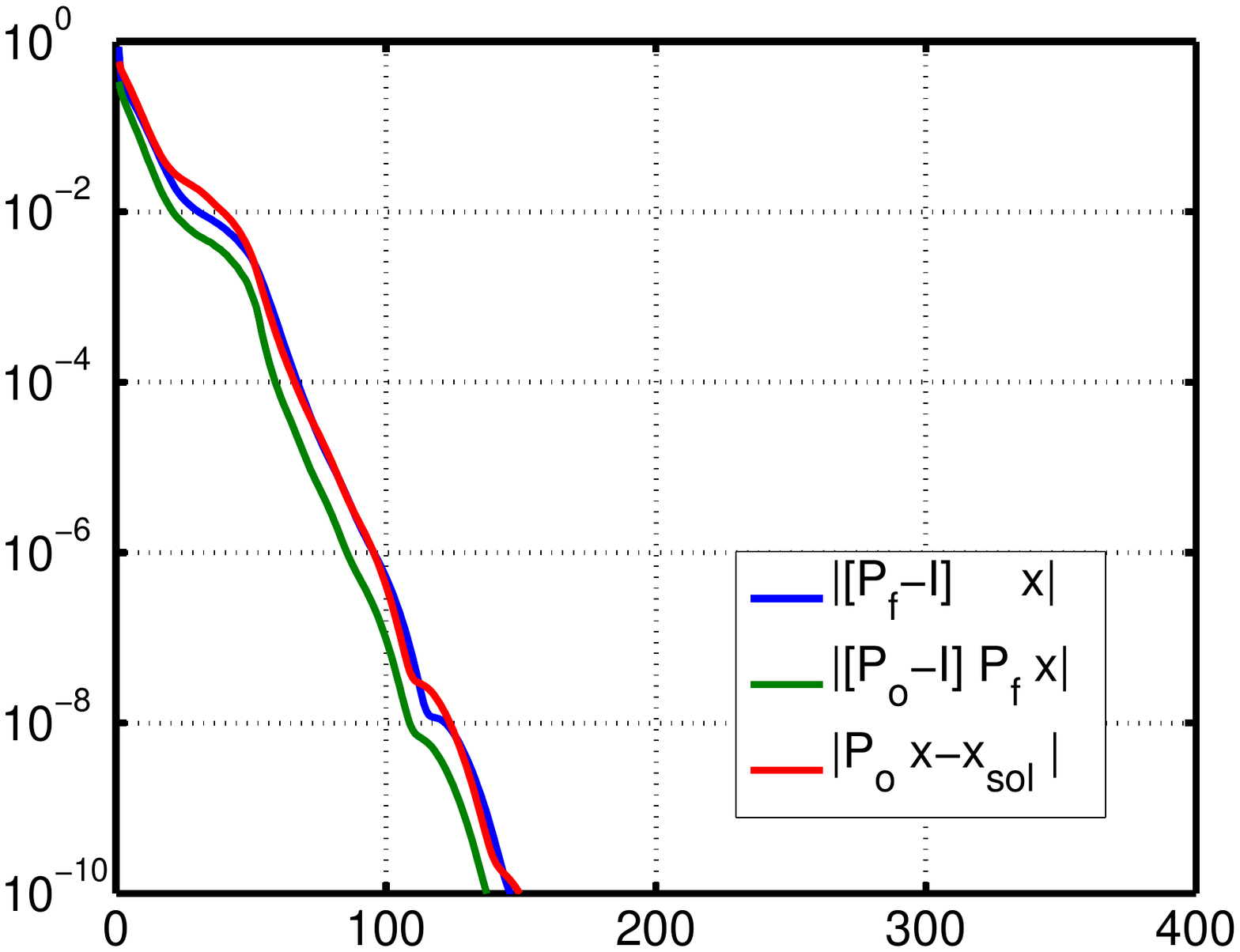}
}
  \caption{
  \label{fig:scaling}
Convergence rate ($\varepsilon_F$, $\varepsilon_Q$, $\varepsilon_0$ vs
number of iteration $\ell$) for (top) regular reconstruction. (bottom)
using augmented projection ($m=16$ and step size $x_1-x_2=3$)
}
\end{figure}

\begin{figure}[hbtp]
  \begin{center}
    \includegraphics[width=0.25\textwidth]{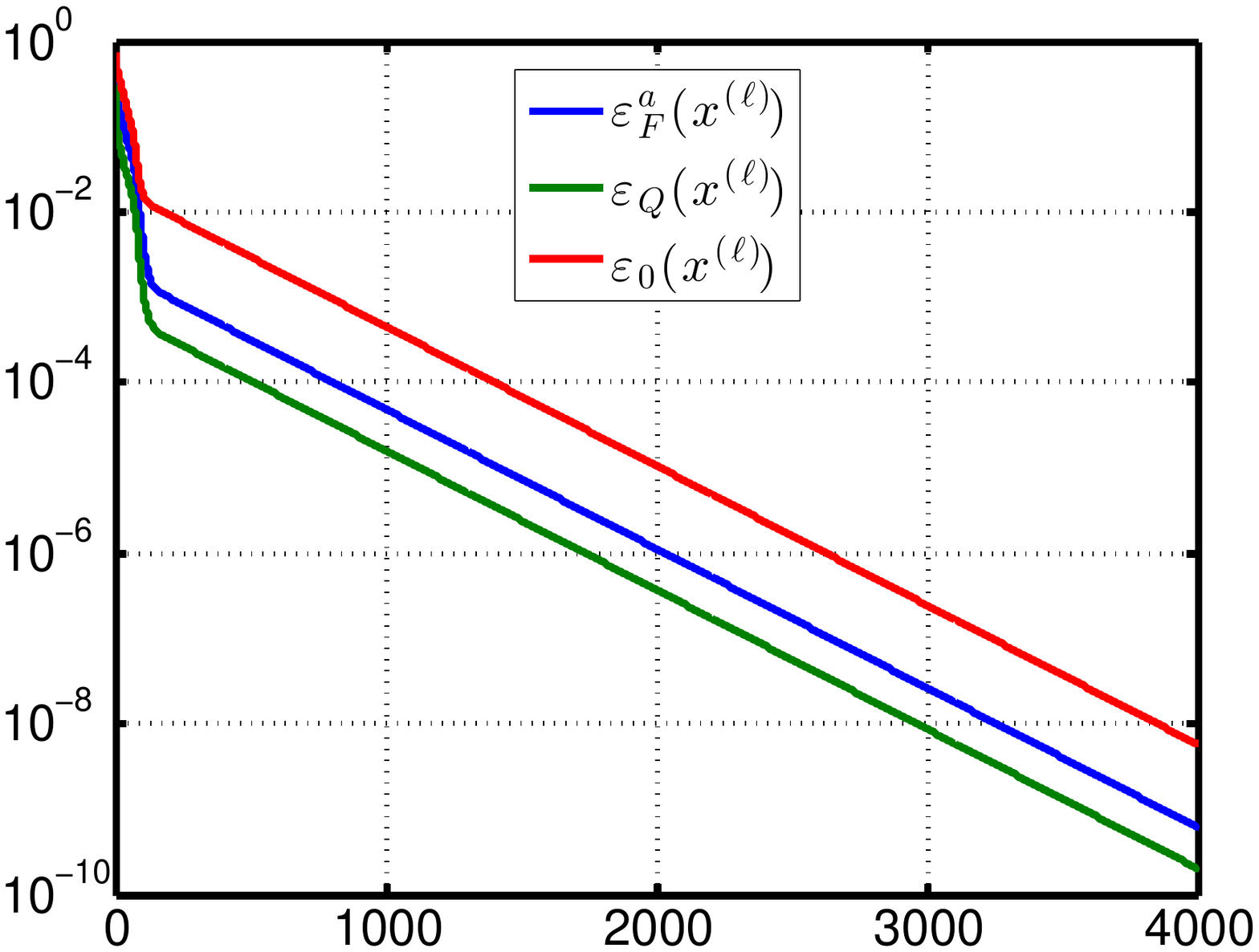}
    \includegraphics[width=0.25\textwidth]{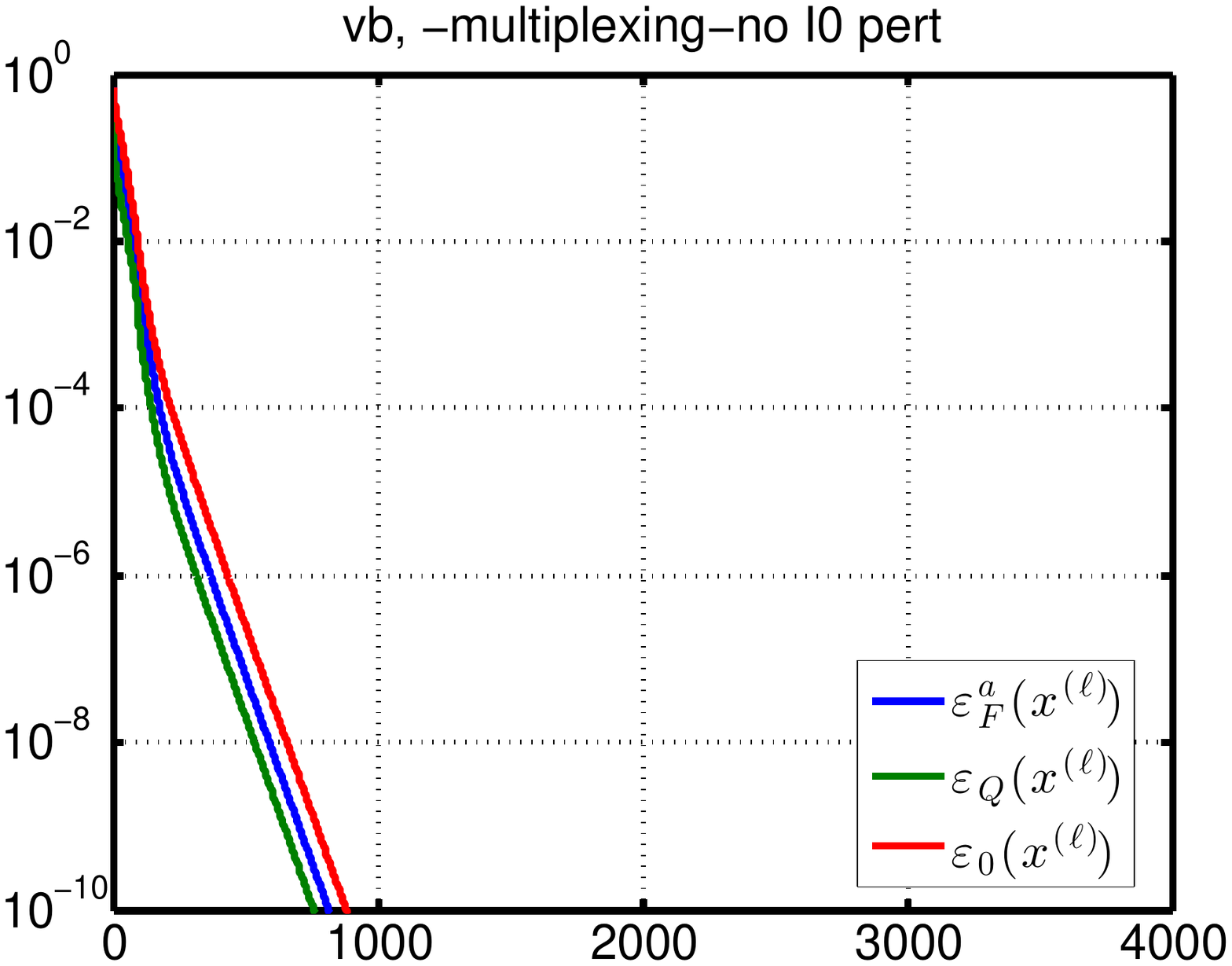}\\
    \includegraphics[width=0.25\textwidth]{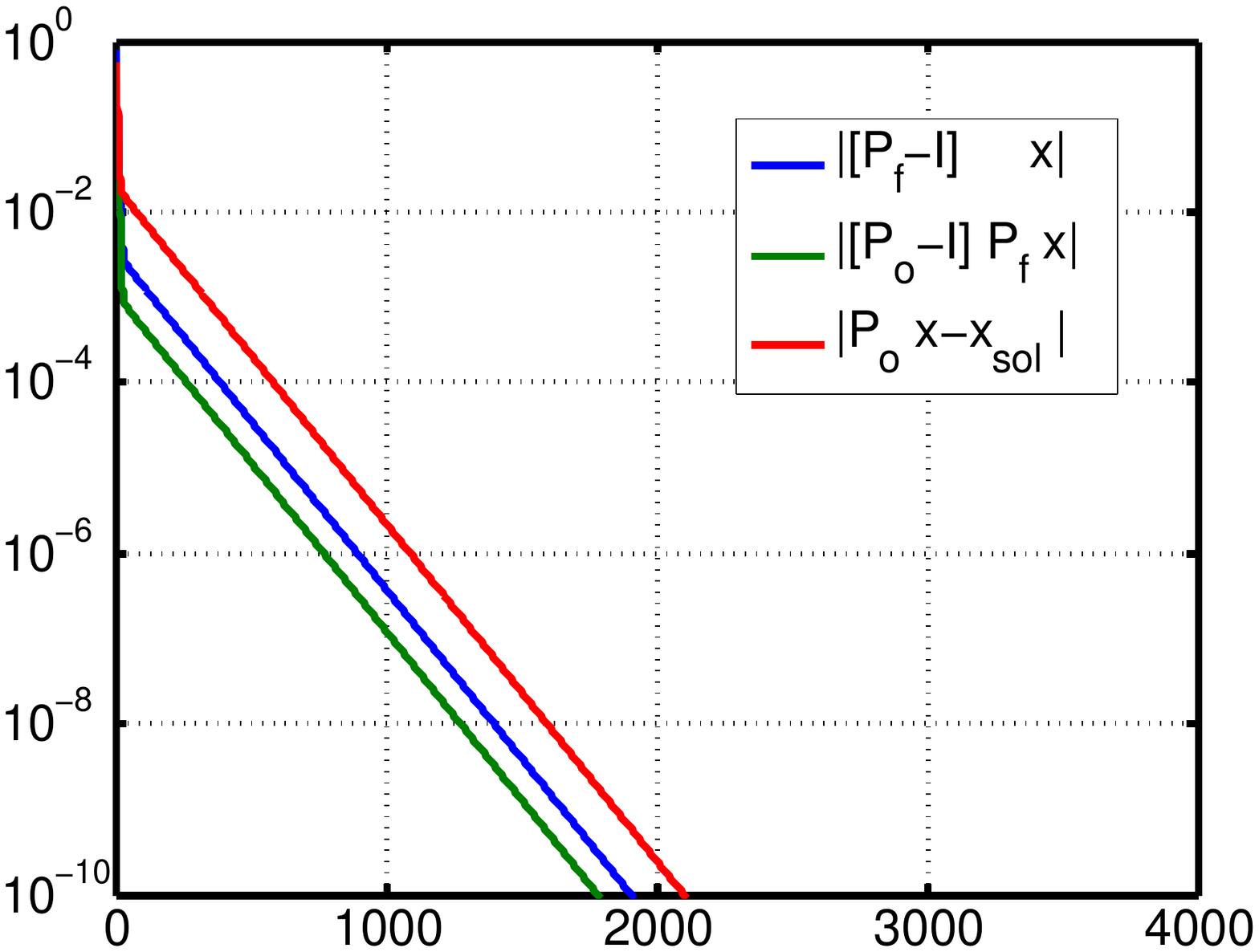}
    \includegraphics[width=0.25\textwidth]{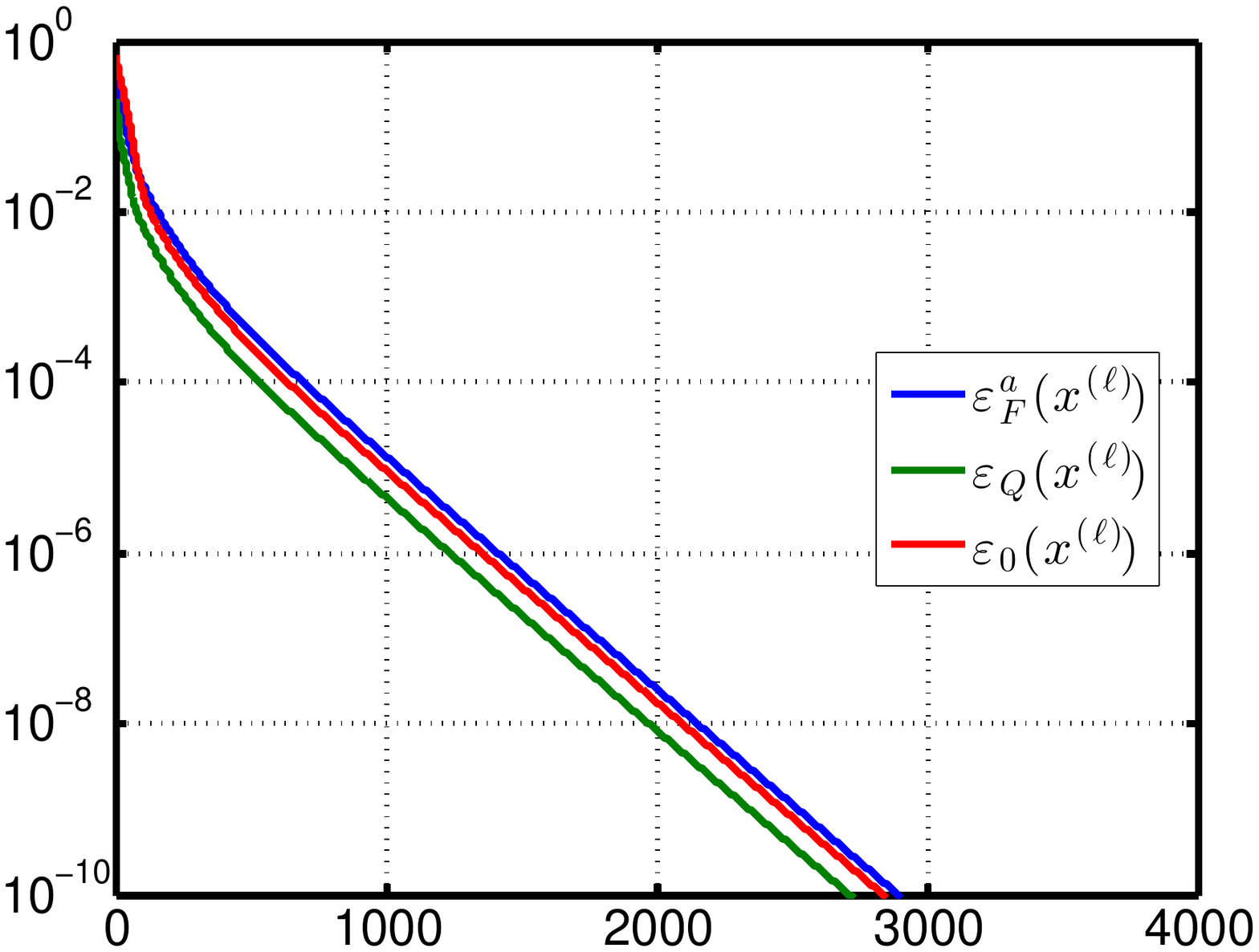}
  \end{center}
  \caption{ Convergence rate with incoherent illumination of 4 beams,
    separated by $3\times$ the probe width (FWHM) using standard
    projection algorithms (top left), with intermediate phase
    optimization (top-right), phase and amplitude(bottom-left), and
    phase and amplitude with initial amplitude error of 20\% (bottom
    right), frame width $16\times16$, $16\times 16$ frames, step size
    $3.5$ pixels close packing with $\pm1$ pixel known random
    perturbations.
  \label{fig:multiplexing}
}
\end{figure}

\begin{figure}[hbtp]
  \begin{center}
    \includegraphics[width=0.4\textwidth]{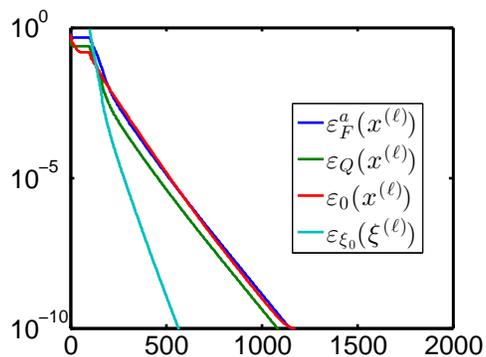}
  \end{center}
  \caption{Reconstruction with position errors using the method described in
    section \ref{sec:positions}, where  $\varepsilon_{\xi_0}=\|\xi-\xi_0\|/\|\xi_0\|$, and the
    perturbations in position are randomly distributed with $\langle
    \xi_0 \rangle= \tfrac 1 k \sum_i\|\xi_i \|=2.5$ resolution
    elements. (number of frames: 16$\times$16, frame dimensions
    $32\times32$, step size: 3.5 pixels, hexagonal packing with known
    random perturbations of $\pm1$ pixels and unknown $\xi$ random
    perturbations ).
  \label{fig:positions}
}
\end{figure}

\begin{figure}[hbtp]
  \begin{center}
    \includegraphics[width=0.3\textwidth]{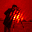}
    \includegraphics[width=0.3\textwidth]{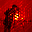}
  \end{center}
  \caption{Two measured intensities with additive background (SNR=0.5).
In a separate test the diffraction data was buried by the background (
(in other figures not included background was $10^6\times$  the signal).
  \label{fig:bkgdata}
}
\end{figure}

\begin{figure}[hbtp]
  \begin{center}
    \includegraphics[width=0.25\textwidth]{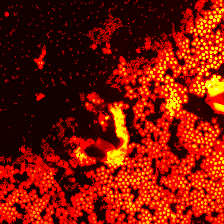}
    \includegraphics[width=0.25\textwidth]{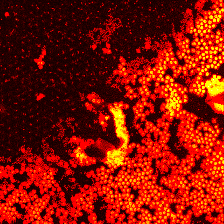}\\
    \includegraphics[width=0.25\textwidth]{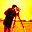}
    \includegraphics[width=0.25\textwidth]{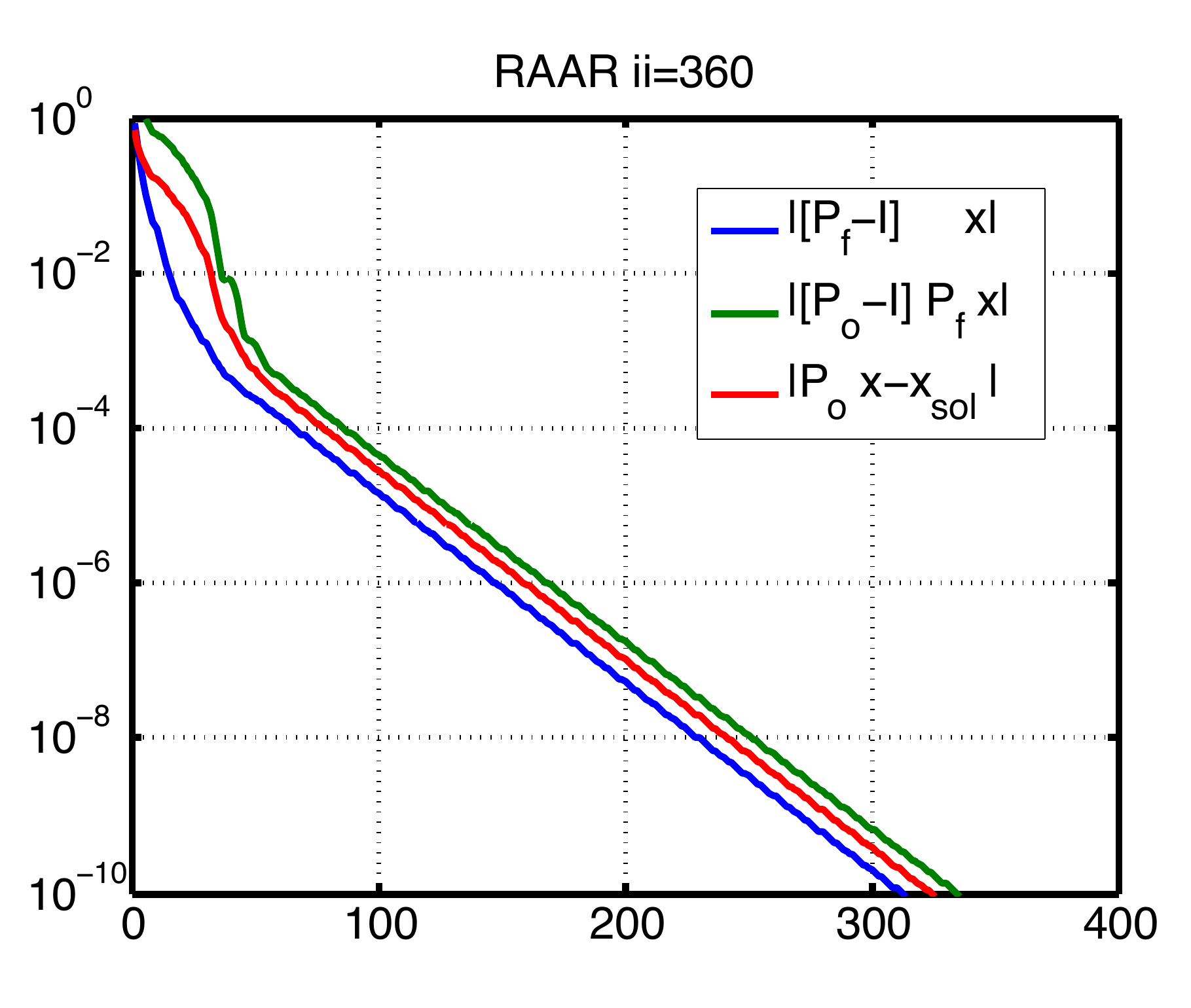}
  \end{center}
  \caption{ (top) reconstructed image with background optimization (left) and without (right). 
     { The figure on the left looks identical to the exact solution.  }
  (bottom) reconstructed background
    (left), convergence behavior(right).
        \label{fig:bkg}}
\end{figure}

\begin{figure}
\begin{minipage}[t]{0.45\linewidth}
\centering
    \includegraphics[width=.5\linewidth]{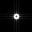}
\caption{
Fourier transform of the probe $w$ used to simulate the experiment
}\label{fig:Fprobe}
\end{minipage}
\hspace{0.5cm}
\begin{minipage}[t]{0.45\linewidth}
\centering
\frame{    \includegraphics[width=.5\linewidth]{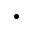}}
\caption{
Map of $\frac 1 {1+\sigma^2}$. The $\sigma$ are $\infty$ on dark pixels and $0$ on white pixels.
}\label{fig:bstop}
\end{minipage}\\
\begin{minipage}[p]{0.95\linewidth}
\centering
    \includegraphics[width=.5\linewidth,clip]{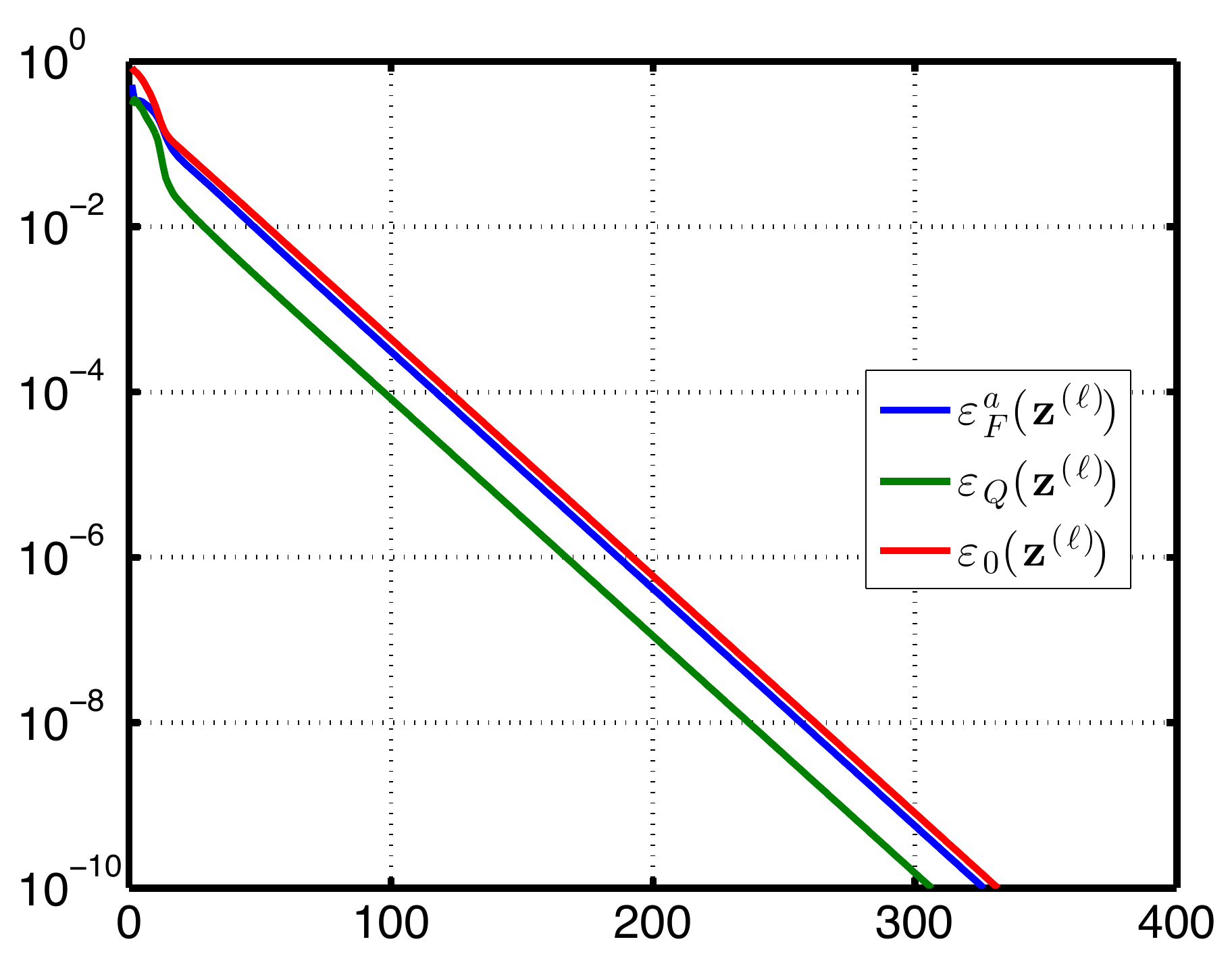}
\caption{Convergence behavior with missing data.
Frame size: $32\times 32$, number of frames: $16\times16$, step size: $3.5$ pixels \label{fig:bstop1}
}\end{minipage}
\end{figure}

\bibliographystyle{unsrt}
\bibliography{ptyco}
\end{document}